\documentstyle[aps,prl]{revtex}
\input epsf.sty
\begin{document}
\title{Effects of Long-Range Correlations in Random-Mass
Dirac Fermions}
\author{Koujin Takeda\cite{Takeda}}
\address{Institute for Cosmic Ray Research, University of Tokyo, Kashiwa, 
Chiba, 277-8582 Japan
}
\author{Ikuo Ichinose\cite{Ichinose}}
\address{Department of Electrical and Computer Engineering, 
Nagoya Institute of Technology, Gokiso, Showa-ku, 
Nagoya, 466-8555 Japan
}

\maketitle
\begin{abstract}
In the previous paper, 
we studied the random-mass Dirac fermion in one
dimension by using the transfer-matrix methods.
We furthermore employed the imaginary vector potential methods 
for calculating the localization lengths.
Especially we investigated effects of the nonlocal but short-range 
correlations of the random mass.
In this paper, we shall study effects of the long-range correlations of 
the random mass especially on the delocalization transition and 
singular behaviours at the band center.
We calculate localization lengths and density of states 
for various nonlocally correlated random mass.
We show that there occurs a ``phase transition" as  the 
correlation length of the random Dirac mass is varied.
The Thouless formula, which relates the density of states and the 
localization lengths, plays an important role in our investigation.

\end{abstract}

\section{Introduction}
In the previous papers \cite{PPs,KI1}, we studied the random-mass 
Dirac fermions 
in one dimension by using the transfer-matrix methods (TMM)
and imaginary-vector potential methods (IVPM).
We calculated the density of states, the typical and mean localization
lengths, and the multifractal scalings
as a function of the energy and correlation length of the
random mass.
The results are in good agreement with the available analytical
calculations \cite{SUSY,IK}.
We also obtained the relation between the correlation length
of the random mass and the typical localization length
for the short-range correlations \cite{KI1}.

In this paper, we shall study effects of the long-range correlations,
especially on the delocalization transition and singular behaviours
at the band center.
This problem is interesting for various reasons.
Very recently finite mobility edges are 
observed in the one-dimensional Anderson model with the long-range correlated
potentials though it is widely believed that
(almost) all states are localized in spatial one dimension 
\cite{lyra,izrailev}.
Similar phenomena are observed also in the aperiodic Kronig-Penney
model \cite{IzKrUl}.
In contrast to the above models, the system of
the random-mass Dirac fermions belongs to the universality class of the 
{\em chiral}
orthogonal ensemble and therefore the extended states exist at the 
band center though the others are localized for the white-noise
random mass.
Long-range correlations of the disorder may change this feature.
This problem has not been addressed in detail so far.
As the model is closely related with the random-bond XY model, the random 
Ising model, etc., the results are interesting both theoretically and 
experimentally.

This paper is organized as follows.
The model and numerical methods are explained in the previous paper \cite{KI1}.
In Sec.2, we shall explain how to make a long-range correlated random
Dirac mass numerically.
We shall consider two types of the telegraphic random mass, i.e.,
in the first one magnitudes of the random mass
are long-range correlated random variables
(LRCRV) with fixed interval distances between kinks
whereas in the second one interval distances between kinks are 
LRCRV with a fixed random mass magnitude.
In the present study, we employ the first one in the above for
it generates various correlations of the Dirac mass more easily
than the second one.
In Sec.3, numerical results are given.
We calculate the localization lengths and the density of states (DOS)
directly by using the TMM and IVPM
though in most of studies of one-dimensional random models
elaborated techniques like the Hamiltonian mapping,
the renormalization group, etc. are used for calculating the Lyapunov exponent.
For the white-noise and short-range correlated random mass, the
DOS $\rho(E)$ diverges as $\rho(E) \propto {1\over E|\log E|^3}$
where $E$ is the energy measured from the band center.
Similarly the typical localization length $\xi(E)\propto |\log E|$.
This means that there are extended states at the band center.
We study how these behaviours change by the existence of 
the long-range correlations of the Dirac mass.
We show that there is a kind of ``phase transition" as the correlation
length is varied.
Section 4 is devoted to conclusion.

\section{Models and Long-range correlated random mass}

Hamiltonian of the random-mass Dirac fermion is given by
\begin{eqnarray}
{\cal H}&=&\int dx \psi^\dagger h\psi, \nonumber   \\ 
h&=&-i\sigma^z \partial_x +m(x)\sigma^y,
\label{Hdirac}
\end{eqnarray}
where $\vec{\sigma}$ are the Pauli matrices and $m(x)$ is the
telegraphic random mass. 
It is known that almost all energy eigenstates of the Hamiltonian
(\ref{Hdirac}) away from the band center are localized for short-range
correlated $m(x)$ \cite{KI1}.
There we calculated localization length of each state by introducing
an imaginary-vector potential (IVP) $\tilde{g}$ as suggested by the 
work of Hatano and Nelson \cite{HN}.
In an IVP, all localized states tend to extend
and a ``critical" value of IVP $\tilde{g}_c$ at which a state 
becomes extended, ``determines" its localization length at 
$\tilde{g}=0$ \cite{KI1}.

In this paper we shall consider the long-range correlated
configurations of $m(x)$ and study delocalized states by
the TMM and IVPM.
Numerical methods which generate potentials correlated by the power law was
recently invented by Izrailev {\it et.al.} and Herbut
\cite{izrailev,herbut}.
We briefly review it for we shall use it for the present studies.

Let us consider a spatial lattice and random potential $\epsilon(n)$
sitting on the sites which has correlation $\chi(n)$,
\begin{equation}
  [\ \epsilon (m) \; \epsilon (n) \ ]_{\rm ens}= C\; \chi{(|m-n|)},
\label{disordercor2}
\end{equation}
where $n$ and $m$ are site indices, and $C$ is a constant.
Here we suppose that $\chi(n)$ is given and define 
$\tilde{c}(k)$ from $\chi(n)$ as follows,
\begin{equation}
  \tilde{c}(k) = \Big(\tilde{\chi}(k)\Big)^{1/2},
\end{equation}
where $\tilde{\chi}(k)$ is the Fourier transformation of $\chi(n)$;
\begin{equation}
\tilde{\chi}(k)= \sum^{\infty}_{n=-\infty} \chi(n)e^{ikn}.
\end{equation}
We also introduce another random site potential $r(n)$ 
which is distributed uniformly in the range $[-1,1]$ with
the delta-function white-noise correlation;
\begin{equation} 
 [\ r(m) \; r(n)\ ]_{\rm ens} \propto \delta_{mn}.
\end{equation}
 Then we can construct random potential $\epsilon(n)$ as follows
 by using the white-noise 
 potential $r(m)$ and the inverse Fourier transformation of $\tilde{c}(k)$,
\begin{equation}
 \epsilon (n) = C^{'} \sum^{\infty}_{m=-\infty} r (m)\; c(m-n),
\label{eps}
\end{equation} 
where $C^{'}$ is another constant.
However, we cannot perform the infinite summation in (\ref{eps})
in the practical numerical calculation, and therefore
we replace it with the finite summation as follows,
\begin{equation}
\label{eq:construct}
 \epsilon (n) = C^{'} \sum^{k/2}_{m=-k/2} r (m)\; c(m-n).
\end{equation} 
 where $k$ is a finite number which is 
 much larger than the number of sites in the system.
We use the periodic boundary condition for the numerical calculation
in subsequent sections and
set $k \simeq 10^{4}+\mbox{(number of sites)}$.
 
 We can easily verify that the random variable $\epsilon(n)$ in 
(\ref{eq:construct}) satisfies Eq.(\ref{disordercor2}).
 If we choose $c(m)$ as
\begin{equation}
 c (m) = \frac {1}{|m|^{\alpha}},
\end{equation}
with some constant $\alpha$,
then the correlator of the random potential becomes
\begin{equation}
[\ \epsilon (m) \; \epsilon (n) \ ]_{\rm ens}  \sim \frac{1}
 {|m-n|^{2 \alpha -1} }.
 \label{ep-correlation}
\end{equation}
 With this procedure, 
 we can generate power-law correlated random 
 potential. Similarly, we can generate exponentially correlated 
 random potential if we choose the modified Bessel function
 $K_{0}$ as $c(m)$.

In the practical calculation,
there is a subtle point in the normalization of the random potential
$\epsilon(n)$ in Eq.(\ref{ep-correlation})
when we vary the system size $L$ \cite{comment}.
We determine the normalization of $\epsilon(m)$ (namely the value of 
 $C^{'}$) in such a way that the typical value of $\epsilon(m)$ does not
 change.
Then as a result, the strength of correlation
 does {\em not} change for various system size $L$ if we fix $|m-n|$ in
 Eq.(\ref{ep-correlation}).
 
The above method generates the random potential at each site.
Therefore, when we fix the
 distances between kinks and vary the magnitudes of $m(x)$
for generating random $m(x)$, we can
 directly use the above methods (see Fig.1). 
 We can easily extend TMM  
 to the above type of random mass $|m(x)|$.
 Numerical studies of this system will be reported in the following section.

\begin{figure}
\label{fig:example1}
\begin{center}
\unitlength=1cm

\begin{picture}(20,3)
\centerline{
\epsfysize=4cm
\epsfbox{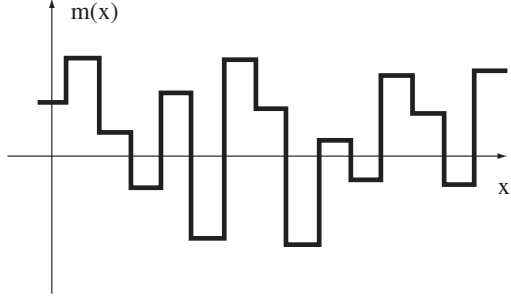}
}
\end{picture}
\vspace{-3mm}
\caption{An example of $m(x)$}
\vspace{0cm}
\end{center}
\end{figure}

 On the other hand, we can also use the above method for generating
random distances between kinks with fixed value of $|m(x)|$.
This is the subject in our previous papers \cite{PPs,KI1}.

\section{Long-range correlated disorders}

We focus on the system with exponential and power-law correlated random mass
in this section.
Each correlation is parameterized as follows;
\begin{eqnarray}
&& [\ m(x) \; m(y)\ ]_{\rm ens} = \frac{g}{2 \lambda} \; 
\exp (\frac{-|x-y|}{\lambda}),\;\; \mbox{for exp-decay} \nonumber \\
&& [\ m(x) \; m(y)\ ]_{\rm ens}={C \over |x-y|^{\alpha_{pw}}},
\;\; \mbox{for the power-law decay}
\label{mmens}
\end{eqnarray}
\begin{figure}
\label{fig:mshape}
\begin{center}
\unitlength=1cm
\begin{picture}(10,3)
\put(-8,-0.3){
\centerline{
\epsfysize=3.7cm
\epsfbox{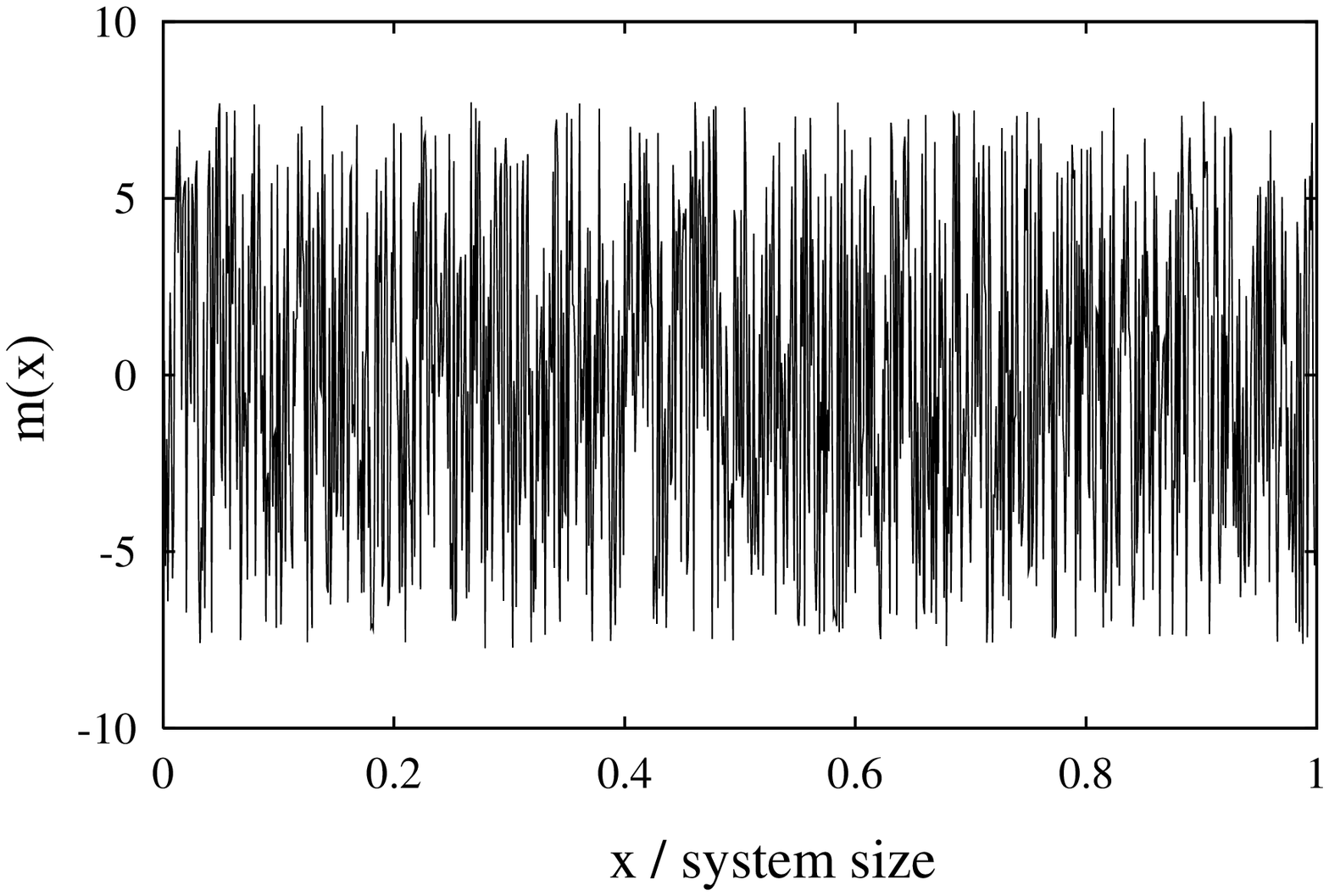}
}}
\put(-1,-0.3){
\centerline{
\epsfysize=3.7cm
\epsfbox{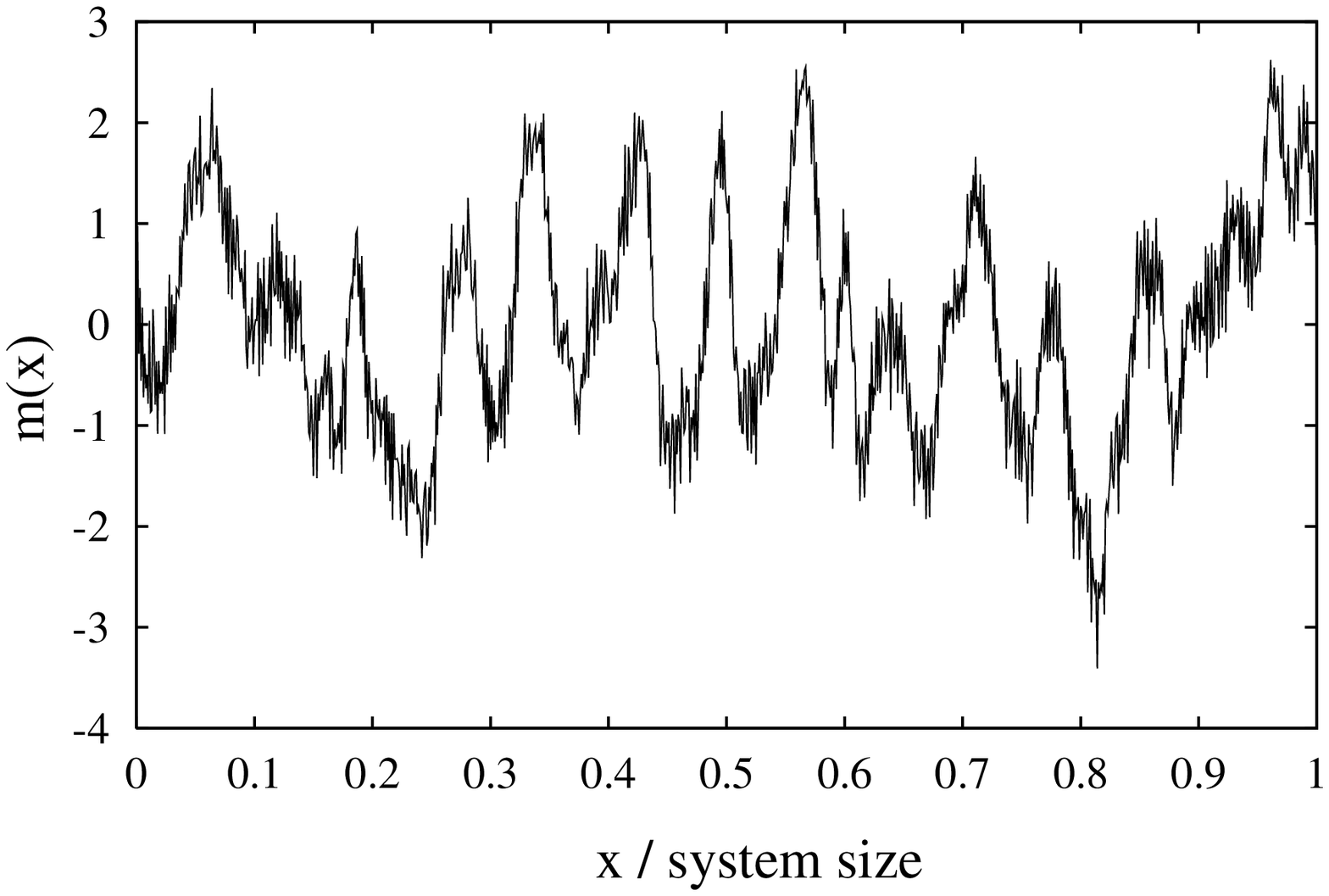}
}}
\put(-8,-4){
\centerline{
\epsfysize=3.7cm
\epsfbox{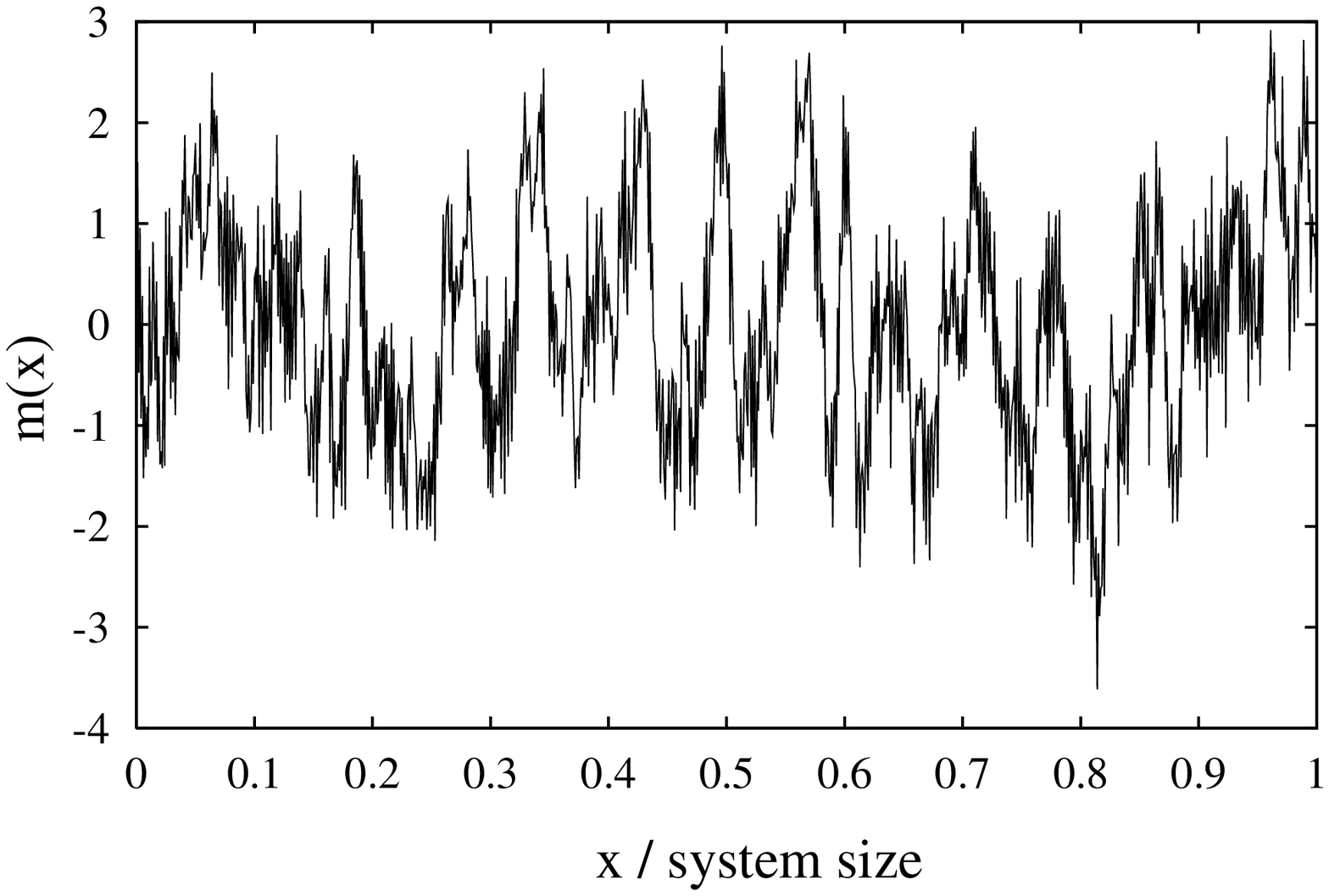}
}}
\end{picture}
\vspace{45mm}
\caption{An example of $m(x)$ in the case of white-noise disorder(top left),
 exponentially correlated(top right) and power-law correlation(bottom): 
 We set 2000 kinks in these systems. By the numerical analysis, 
 correlation length (per system size) $\lambda$ is calculated
 as $0.16$ in exponentially correlated case and $\alpha_{pw}$ is
 $0.68$ in power-law case.}
\end{center}
\end{figure}

where $g, \lambda, C$ and $\alpha_{pw}$ are constants.
Dimensions of the parameters are $[\lambda]=M^{-1}, \; [g]=M^1,\;
[C]=M^{2-\alpha_{pw}}$, respectively.
As we explained in Sec.2, we consider
the random mass of random magnitude with fixed distances between
kinks.
Prototype of the configuration is given in Fig.2.
In this case $m(x)$ is generated rather similarly to
$\epsilon(n)$ in Sec.2 though the continuum space is considered here
instead of the spatial lattice.
As in the previous paper \cite{KI1}, we use the 
IVPM and TMM in order to calculate the localization
lengths.

We first calculate the correlation $[ m(x) m(0) ]_{\rm ens}$
numerically which is expected to exhibit the exponential and power-law decay,
respectively.
The result is shown in Fig.3.
The correlator actually shows the expected behaviour in each case, 
though the parameters
of the correlation are slightly smaller than the analytical values, which are 
$\lambda$ in exponential-decay case and $\alpha_{pw}$ in power-law decay case,
respectively.
We think that this is due to the finiteness of numbers of kinks
and/or {\em finite} Fourier transformation in the scheme which
we explained in the previous section. 
(The values of $\alpha_{pw}$, $\lambda$ 
and the parameters in Tables 1 and 2 are calculated directly from the
two-point correlation function of the random mass obtained numerically.)

\begin{figure}
\label{fig:powerlaw}
\begin{center}
\unitlength=1cm

\begin{picture}(10,4.8)
\put(-8,1){
\centerline{
\epsfysize=4cm
\epsfbox{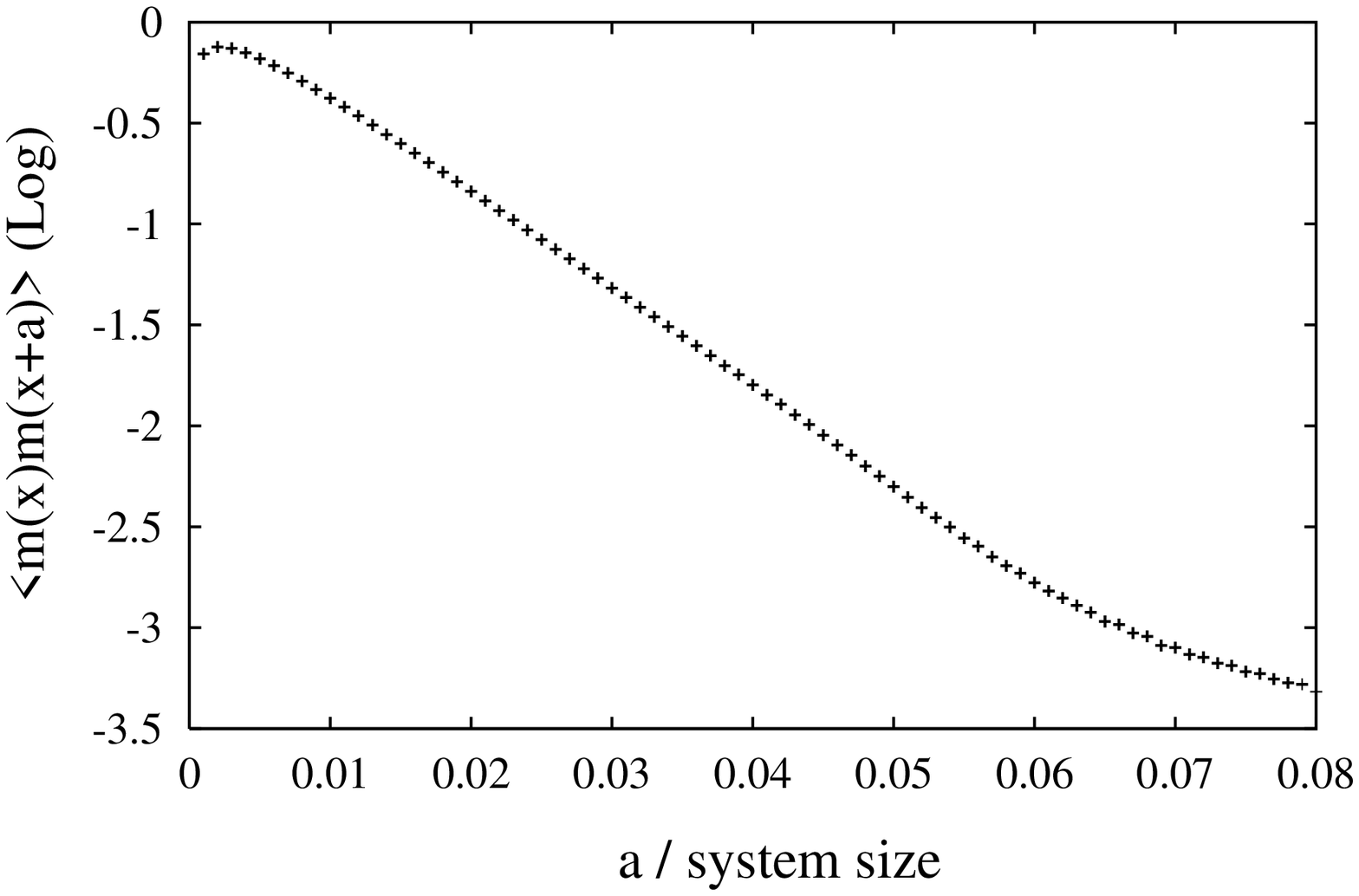}
}}
\put(-1,1){
\centerline{
\epsfysize=4cm
\epsfbox{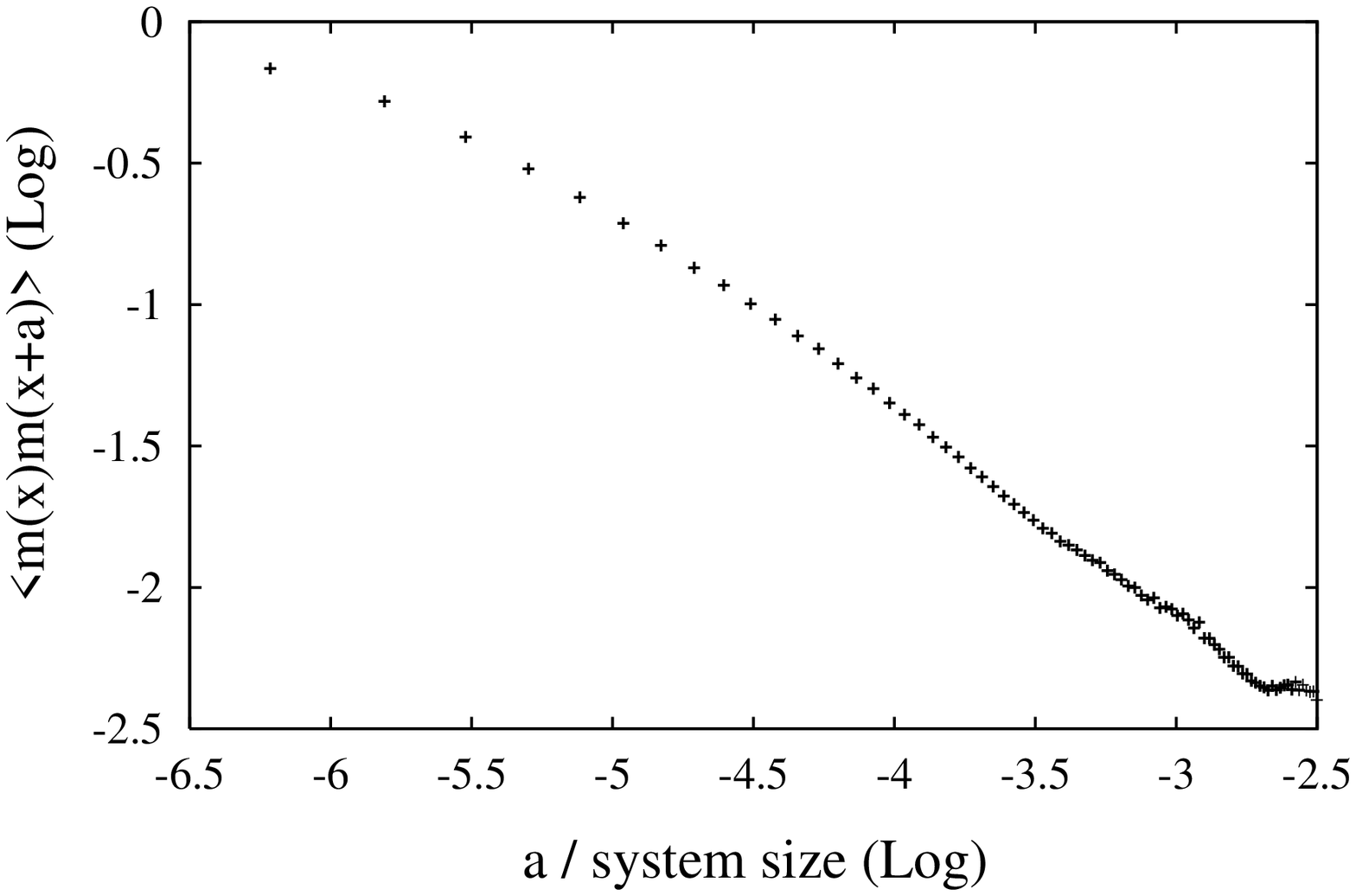}
}}
\end{picture}

\vspace{-0.8cm}

\caption{Correlation of $m(x)$ in the case of exponentially
 correlation(left) and power-law correlation(right): 
 Each case corresponds to random $m(x)$ shown in Fig.2. In these cases, 
 we can consider that randomness has exponential or power-law
 correlation for about 10 percent of system size.
 We fix the parameters in Table 1 (exponential case) and 2 (power-law
 case) using this numerical result. (In order to fix these parameters, we used the
 data points in the range $0.003<a<0.015$ for exponential case
 ($0.01<a<0.05$ for $g\lambda=12$) and $0.01<a<0.1$
 for power-law case.)}
\vspace{3mm}
\end{center}
\end{figure}

Next we show the energy dependence of typical localization length
in the system with the white-noise random mass.
(We obtain ``typical" localization length as a function of energy $E$
by averaging localization lengths
 of the eigenstates within a small range of energy $\Delta E$.)
In Fig.4, we show the ratio of the numerical results to the available 
analytical expression which takes the form 
\begin{equation}
\xi(E)\propto |\log(E/2g)|,
\label{LLwhite}
\end{equation}
where the parameter $g$ gives the magnitude of $m(x)$,
\begin{equation}
[\ m(x) \; m(y)\ ]_{\rm ens} = g \; \delta(x-y).
\label{white}
\end{equation}
It is easily seen that the limit $\lambda\rightarrow 0$ in Eq.(\ref{mmens})
corresponds to Eq.(\ref{white}).
As Fig.4 shows, our numerical calculations and the analytical expression
(\ref{LLwhite}) are in good agreement.
This guarantees validity of our methods of calculation, the TMM and IVPM. 
\begin{figure}
\label{fig:whitenoise}
\begin{center}
\unitlength=1cm

\begin{picture}(19,3.7)
\centerline{
\epsfysize=4.5cm
\epsfbox{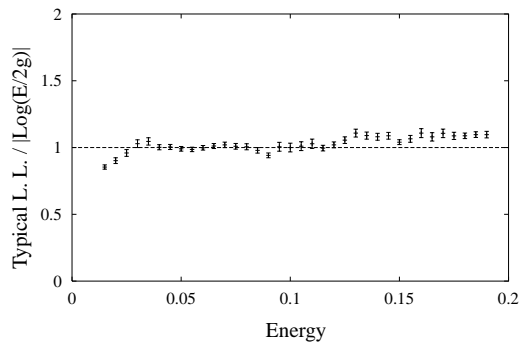}
}
\end{picture}
\vspace{-2mm}
\caption{The energy dependence of localization length in the case
 of white-noise disorder : We set $L$(system size)=50,
 1000 kinks in the system and $g=1$. Ratio is normalized at $E=0.10$
 and data are averaged within small energy slice $\Delta E$, 
 We set $\Delta E=0.02$ here. }
\end{center}
\end{figure}

We calculated the energy dependences of the
localization length in the case of exponentially correlated disorder
in Eq.(\ref{mmens}).
In Fig.5, we show the ratio of numerical calculations to the analytical
expression of the white-noise case $|\log(E/2g)|$.
We vary the correlation length $\lambda$ of $m(x)$.
The ratio is almost constant for small $g\lambda$.
However as $g\lambda$ is getting large, the numerical results obviously
deviate from $|\log(E/2g)|$.  

\begin{figure}
\label{fig:exponential1}
\begin{center}
\unitlength=1cm
\begin{picture}(17,4.8)
\put(-4,1){
\centerline{
\epsfysize=4cm
\epsfbox{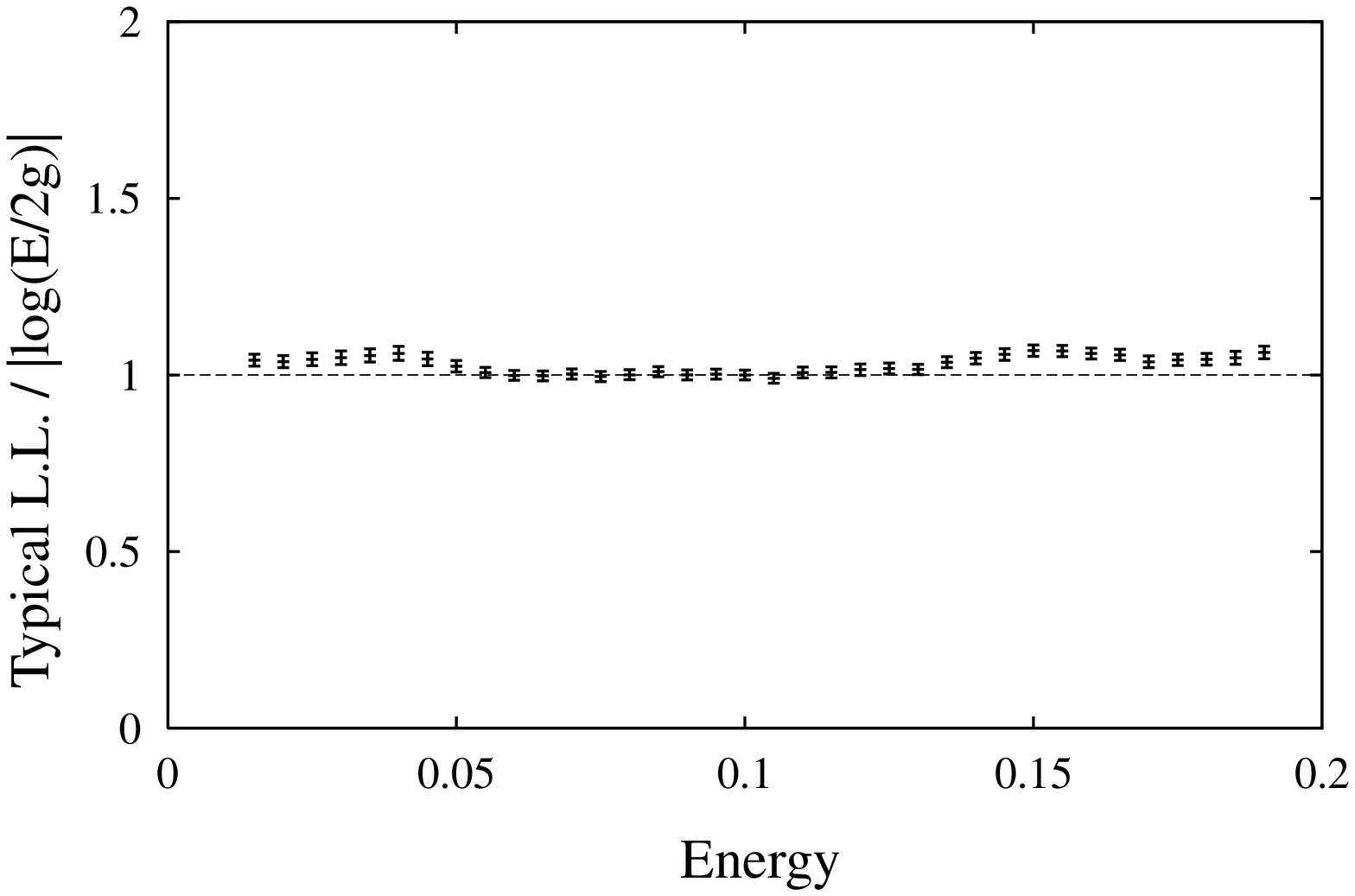}
}}
\put(3,1){
\centerline{
\epsfysize=4cm
\epsfbox{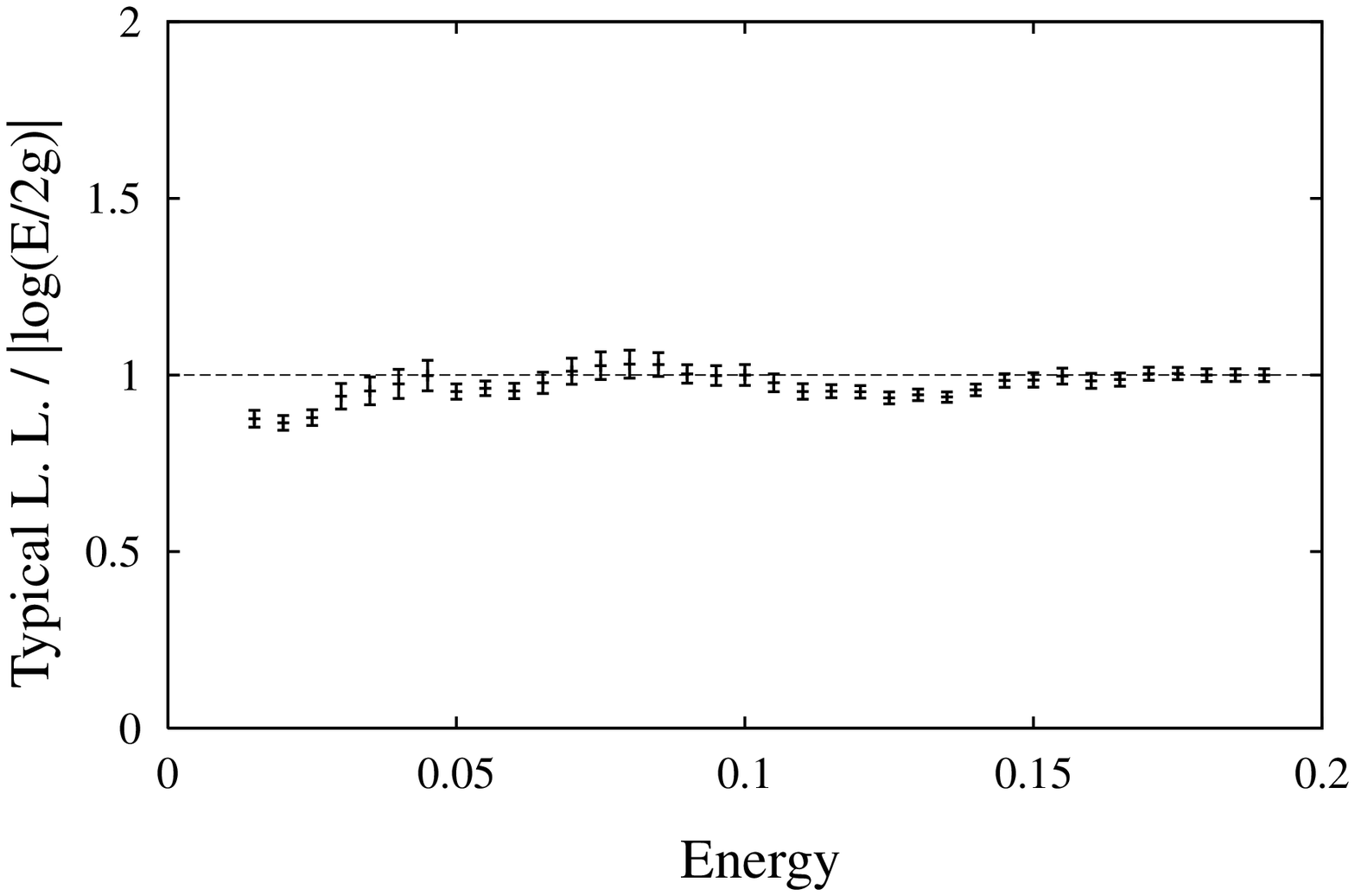}
}}
\put(-4,-3){
\centerline{
\epsfysize=4cm
\epsfbox{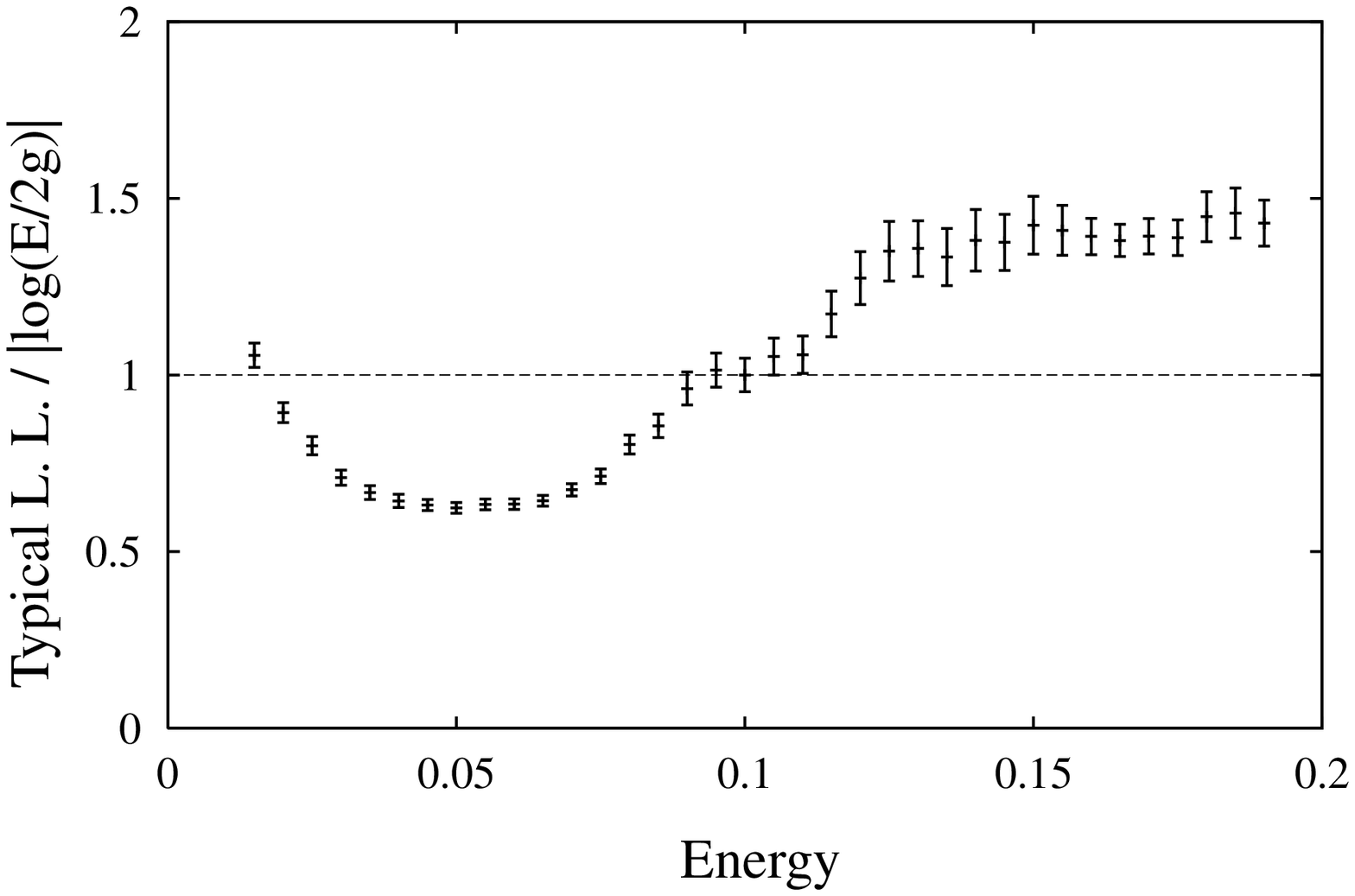}
}}
\end{picture}
\vspace{35mm}
\caption{ The energy dependence of localization length in the case of 
 exponentially correlated random mass: 
 We set $L$(system size)$=50$ and $1000$ kinks in the system. The ratios
 are normalized at $E=0.1$ and data are averaged within
 energy slice $\Delta E=0.02$.
 Values of $\lambda$ and $g$ are given in Table 1.
 The ratio is almost constant for small $g\lambda$, but it deviates from
 constant in the case of large $g\lambda$.}
\end{center}
\vspace{-5mm}
\end{figure}

\begin{center}
\begin{tabular}{ccccc} \cline{2-5} 
 {$ $}&{$ $}&{$\lambda$}&{$g$}&{$g \lambda$} \\ \cline{2-5}
 &\hspace{3mm} top left \hspace{3mm} & $0.28$\hspace{3mm} & $2.9$\hspace{3mm}
 & $0.80$\hspace{7mm}\\  
 { Table 1. Parameters of exponential correlation.} \hspace{10mm}
 &\hspace{3mm} top right \hspace{3mm} & $1.2$\hspace{3mm} & $2.4$\hspace{3mm} 
 & $2.8$\hspace{7mm}\\  
 &\hspace{3mm} bottom\hspace{3mm} & $8.1$\hspace{3mm} & $1.5$\hspace{3mm}
 & $12$\hspace{3mm} \\ \cline{2-5}
 \end{tabular}
\end{center}
\vspace{10mm}

We show the energy dependence of the localization length in the case 
$g\lambda=12$ in Figs.5 (bottom) and 6.
We can conclude that the localization length diverges only at $E=0$
even for very large $g\lambda$. 
This means that exponential correlation of the random mass gives
no significant effect on the {\em Anderson transition}.
Actually
this result can be expected from the study in the previous paper\cite{KI1}.

\begin{figure}
\label{fig:exponential2}
\begin{center}
\unitlength=1cm

\begin{picture}(18,4)
\centerline{
\epsfysize=4.5cm
\epsfbox{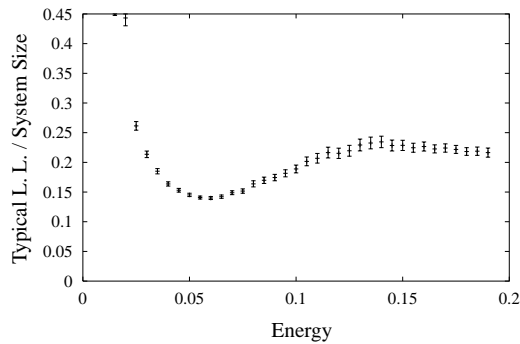}
}
\end{picture}
\vspace{0mm}
\caption{ The energy dependence of localization length in the case of 
 exponentially correlated random mass (especially the case of large
 $g\lambda$): This data corresponds to the one at the bottom in Fig.5. 
 Data are averaged within energy slice $\Delta E=0.02$.
 Divergence of localization length is found only at the point $E=0$.}
\end{center}
\end{figure}

However the result also shows that singular behaviour of the localization
length at $E=0$ for the exponentially correlated $m(x)$
(with large $g\lambda$) may be {\em different from} that of the 
white-noise case.
From the analytical study by using supersymmetry \cite{IK}, 
the localization length is obtained in powers of $g\lambda$ and 
more dominant terms like $|\log (E/2g)|^2$ etc. appear in higher-order
terms of $g\lambda$.

We therefore investigate the behaviour of the localization
length near the band center $E=0$ rather in detail.
In Fig.7 we show the ratio of the numerically obtained localization length to
$|\log(E/2g)|^{\delta}$, where $\delta$ takes the value 1,2,3 and 4.
In some energy regions, the energy dependence is fitted better by the power
 $\delta=2$ or $3$ than $\delta=1$.

\begin{figure}
\begin{center}
\unitlength=1cm
\begin{picture}(10,4.5)
\put(-8,1){
\centerline{
\epsfysize=3.5cm
\epsfbox{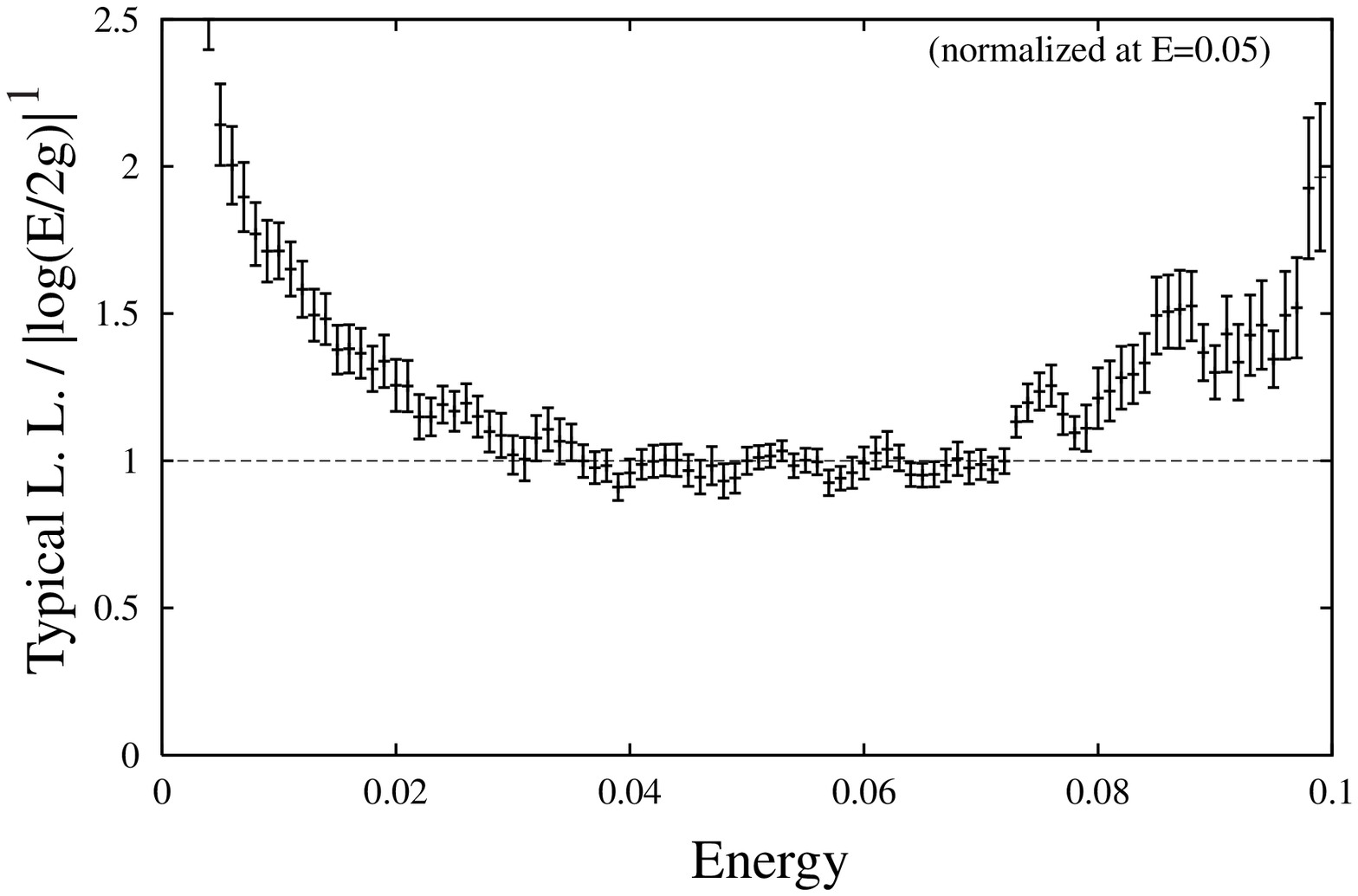}
}}
\put(-1,1){
\centerline{
\epsfysize=3.5cm
\epsfbox{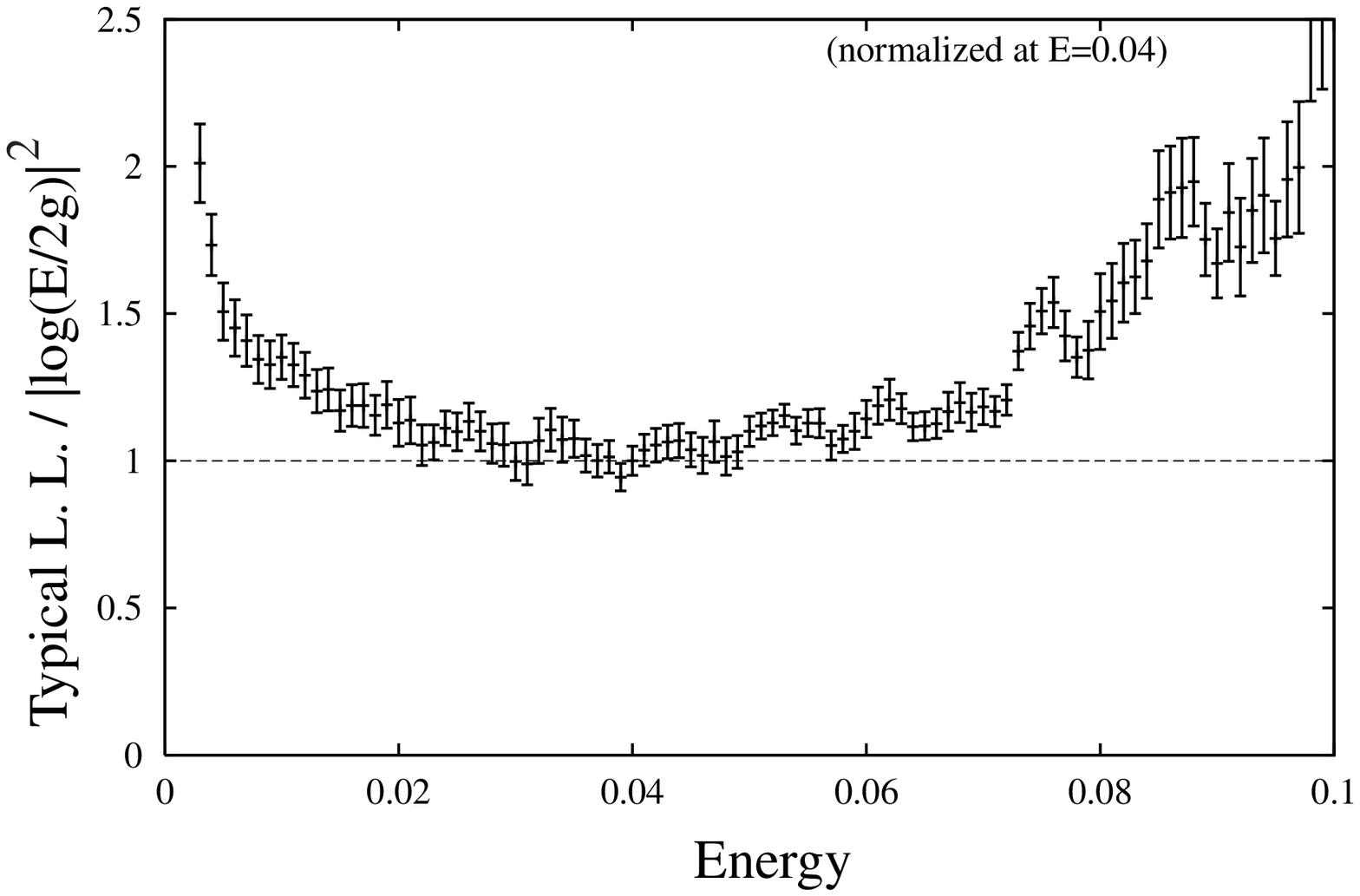}
}}
\put(-8,-2.7){
\centerline{
\epsfysize=3.5cm
\epsfbox{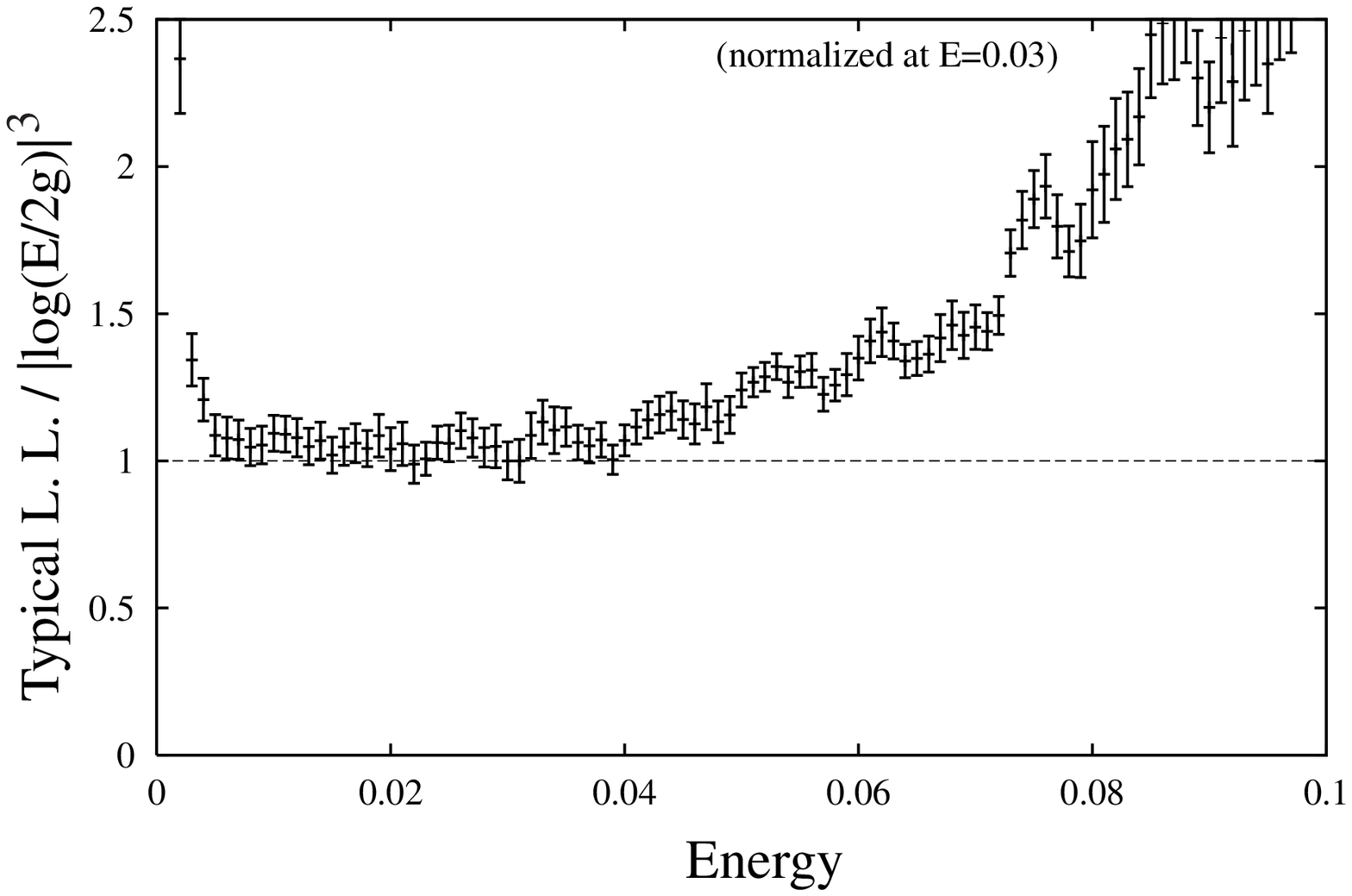}
}}
\put(-1,-2.7){
\centerline{
\epsfysize=3.5cm
\epsfbox{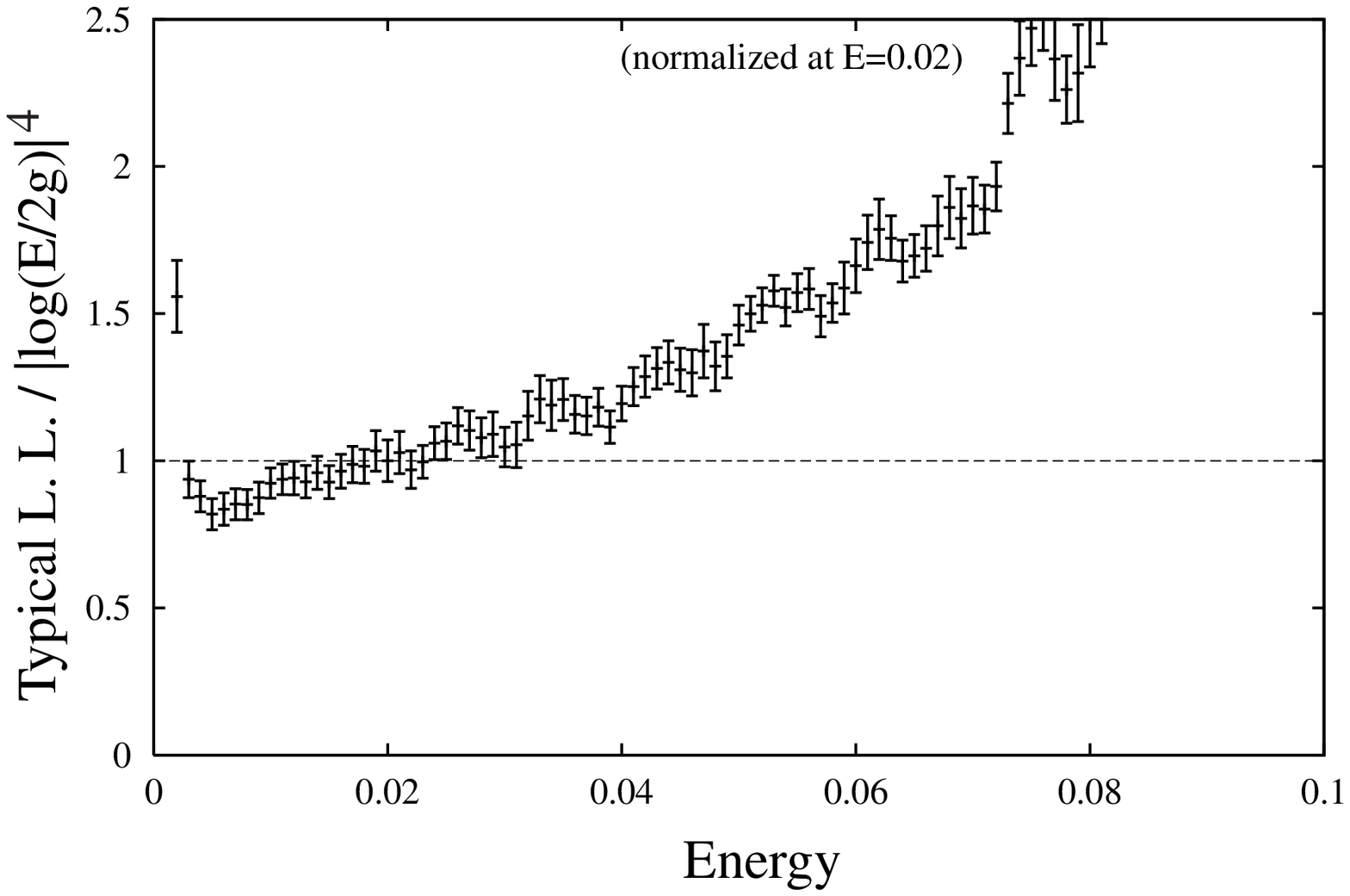}
}}
\end{picture}
\end{center}
\vspace{3cm}
\caption{Ratio of the Localization length to $|\log (E/2g)|^{\delta}$:
Here we set $L$(system size)$=50$, and $\delta$ takes the value 1,2,3
 and 4. Data are averaged within energy slice $\Delta E=0.01$.}
\label{fig:explong}
\end{figure}

We can expect the above result by the analytical study. 
In the previous paper,\cite{IK} the DOS
$\rho(E)$ is obtained analytically as follows for the exponentially 
correlated random mass;
\begin{equation}
\rho(E) = \frac{A_1}{{E\over 2g} |\log(E/2g)|^3}
+\frac{A_2}{{E\over 2g} |\log(E/2g)|^4}+\frac{A_3}
{{E\over 2g} |\log(E/2g)|^5},
\label{DOS}
\end{equation}
where $A_i$ $(i=1,2,3)$ are polynomials of $g\lambda$, 
\begin{eqnarray}
A_1&=&{1\over 2}-{26\over 15}(g\lambda)^2-{4\over 105}(g\lambda)^3
+\cdots, \nonumber \\
A_2&=&-3(g\lambda)-2(g\lambda)^2+\cdots, \nonumber  \\
A_3&=&4(g\lambda)^2 +\cdots.
\end{eqnarray}
In order to verify Eq.(\ref{DOS}),
we calculate the DOS numerically for relatively small $g\lambda(<1)$
and compare the result with Eq.(\ref{DOS}).
From Fig.8,
it is obvious that $\rho(E)$ up to the third-order term of $(g\lambda)$
is in better agreement with the numerical result than the first-order
one.

Using Thouless formula for 1D random hopping tight binding model
\cite{thouless} (whose low-energy field theory is the random-mass 
Dirac fermion),
\begin{equation}
\label{eq:thouless}
\frac{1}{\xi(E)} \propto \int_{0}^{|E|} \log(\frac{E}{E'}) \rho(E') dE',
\end{equation} 
we can show that corresponding to
the second and the third term in (\ref{DOS}) there are contributions to
the localization length like $|\log (E/2g)|^{2}$ and
$|\log (E/2g)|^{3}$ if these terms become dominant in the DOS.
It is also expected that the ratio is not constant for any range of energy
in the case $\delta=4$,
because DOS has no term proportional to $1/E|\log(E/2g)|^{6}$. 

These numerical results and the above discussion indicate that
the nonlocal correlations of the random mass generate 
nontrivial effect on the localization.
One may expect that for large $g\lambda$ the perturbative calculation
in powers of $g\lambda$ breaks down for the DOS and the localization
length and $\rho(E)$ and $\xi(E)$ exhibit different singular behaviours
at $E=0$.
Especially the terms like $|\log E|^\delta$ in $\xi(E)$ may be summed up
and the localization length may behave as 
\begin{equation}
\xi(E) \sim E^{-\beta}
\label{Epower}
\end{equation}
for large $g\lambda$ where $\beta$ is a constant.
In Fig.9 we show the log-log plot of $\xi(E)$ v.s. energy.
The numerical data are obviously on a straight line and therefore
the singular behaviour (\ref{Epower}) is verified.

From the Thouless formula (\ref{eq:thouless}), we also expect the power 
behaviour of the DOS.
Actually it is not so difficult to show that for the DOS parameterized as
\begin{equation}
\rho(E)\sim E^{-\eta}|\log E|^{-\gamma},
\label{rhoE}
\end{equation}
where $\eta$ and $\gamma$ are parameters, the Thouless formula 
(\ref{eq:thouless}) gives the following localization length,
\begin{eqnarray}
\xi(E) &\sim& E^{-1+\eta}|\log E|^{-1+\gamma}, \;\; \mbox{for $\eta\neq 1$},
\nonumber   \\
\xi(E) &\sim& |\log E|^{-2+\gamma}, \;\; \mbox{for $\eta=1$}.
\label{xiE}
\end{eqnarray}
In Fig.10 we show the calculations of the DOS for $g\lambda=12$.
It is obvious that the DOS $\rho(E)$ behaves as
\begin{equation}
\rho(E) \sim E^{-\beta^{'}}.
\label{Epower2}
\end{equation}

We cannot fix the exponent $\gamma$ in Eqs. (\ref{rhoE}) and (\ref{xiE})
from these numerical calculations. 
However, we can conclude that $\eta \neq 1$ in the case $g\lambda=12$. 
This confirms the power-law behaviour of the localization length
near the band center $E=0$, which differs from the behaviour of the
white-noise disorder case.

\begin{figure}
\label{fig:expDOSsmall}
\begin{center}
\unitlength=1cm

\begin{picture}(18,3.8)
\centerline{
\epsfysize=4.3cm
\epsfbox{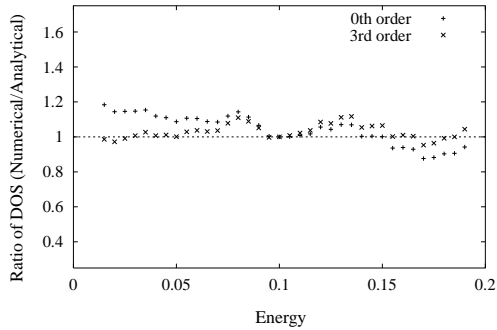}
}
\end{picture}
\vspace{0cm}

\caption{ The energy dependence of DOS in the case of 
 exponentially correlated random mass (especially the case of large
 $g\lambda$): We show the ratio of DOS we calculated numerically to
 the one obtained analytically. Ratio is normalized at $E=0.10$
 and data are averaged within energy slice $\Delta E=0.02$.
 We set $g\lambda=0.80$ here. (This corresponds to the case
 of the top left figure in Fig.5.)
 In order to show the effect of 
 exponential correlation, we use DOS obtained analytically up
 to 0th and 3rd order of $g\lambda$ expansion, respectively.
}
\end{center}
\end{figure}
 
\begin{figure}
\label{fig:expll}
\begin{center}
\unitlength=1cm

\begin{picture}(18,3.2)
\centerline{
\epsfysize=4.3cm
\epsfbox{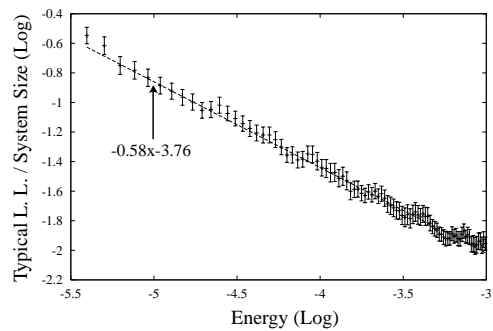}
} 
\end{picture}
\vspace{0cm}
\caption{ The energy dependence of localization length in the case of 
 exponentially correlated random mass (especially the case of large
 $g\lambda$): 
 Data are averaged within energy slice $\Delta E=0.005$.
 We set $g\lambda=12$ here. (This corresponds to the case
 of the bottom figure in Fig.5.) $\chi^{2}$(per freedom) value of this
 fitting is $0.36$.}

\end{center}
\end{figure}
\begin{figure}
\label{fig:expDOSlarge}
\begin{center}
\unitlength=1cm

\begin{picture}(18,4)
\centerline{
\epsfysize=4.3cm
\epsfbox{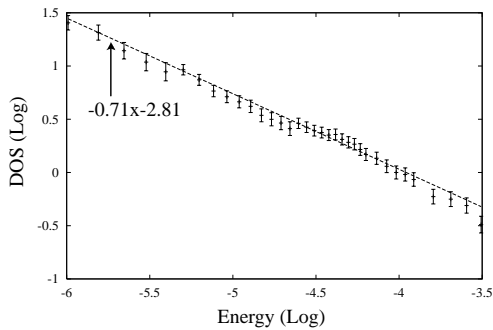}
}
\end{picture}
\vspace{0.3cm}
\caption{ The energy dependence of DOS in the case of 
 exponentially correlated random mass (especially the case of large
 $g\lambda$): 
 Data are averaged within energy slice $\Delta E=0.005$.
 We set $g\lambda=12$ here. (This corresponds to the case
 of the bottom figure in Fig.5.)
  We set energy slice
 $\Delta E=0.002, 0.006$ and $0.01$ for $E<0.005, 0.005<E<0.01$
 and $0.01<E$ respectively.
 The interval of data points are
 $5 \times 10^{-4}, 1 \times 10^{-3}$ and $2.5 \times 10^{-3}$ for
 $E<0.015, 0.015<E<0.02$ and $0.02<E$ respectively. 
 DOS is normalized at $\log E=-4.0$. $\chi^{2}$ value of this fitting
 is $0.64$.}
\end{center}
\end{figure}

Let us turn to the study on the power-law correlated random mass in 
Eq.(\ref{mmens});
$$
[\ m(x) \; m(y)\ ]_{\rm ens} = \frac{C}{|x-y|^{\alpha_{pw}}}.
$$
Here we emphasize significance of the power-law correlation of 
the randomness to the Anderson transition etc.
This is due to the scale-free property of the power-law correlation, and
this type of correlations may not be negligible even though
system size is very large. 
 
\begin{figure}
\label{fig:powerlaw1}
\begin{center}
\unitlength=1cm

\begin{picture}(17,4.5)
\put(-4,1){
\centerline{
\epsfysize=3.8cm
\epsfbox{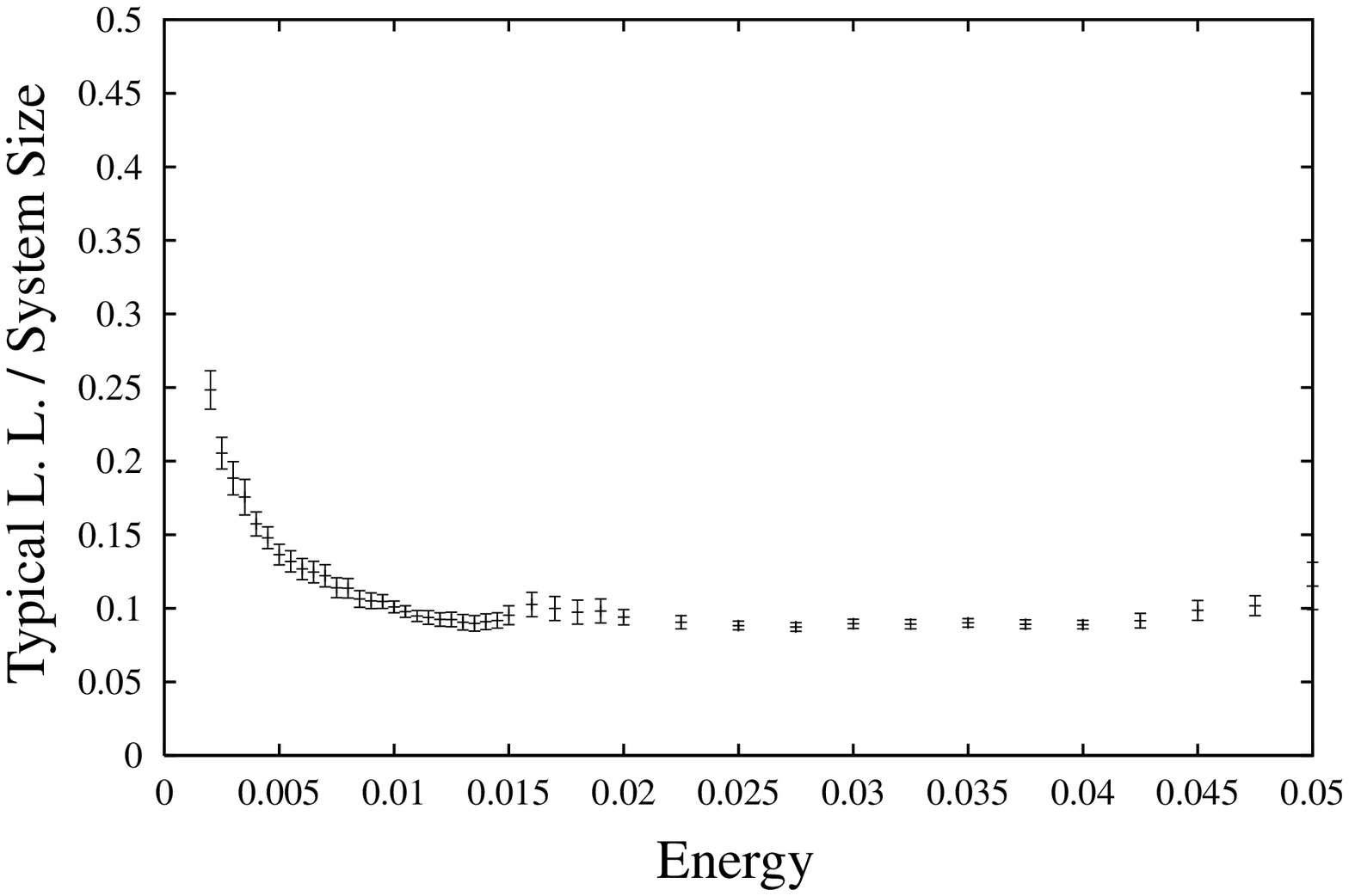}
}}
\put(3,1){
\centerline{
\epsfysize=3.8cm
\epsfbox{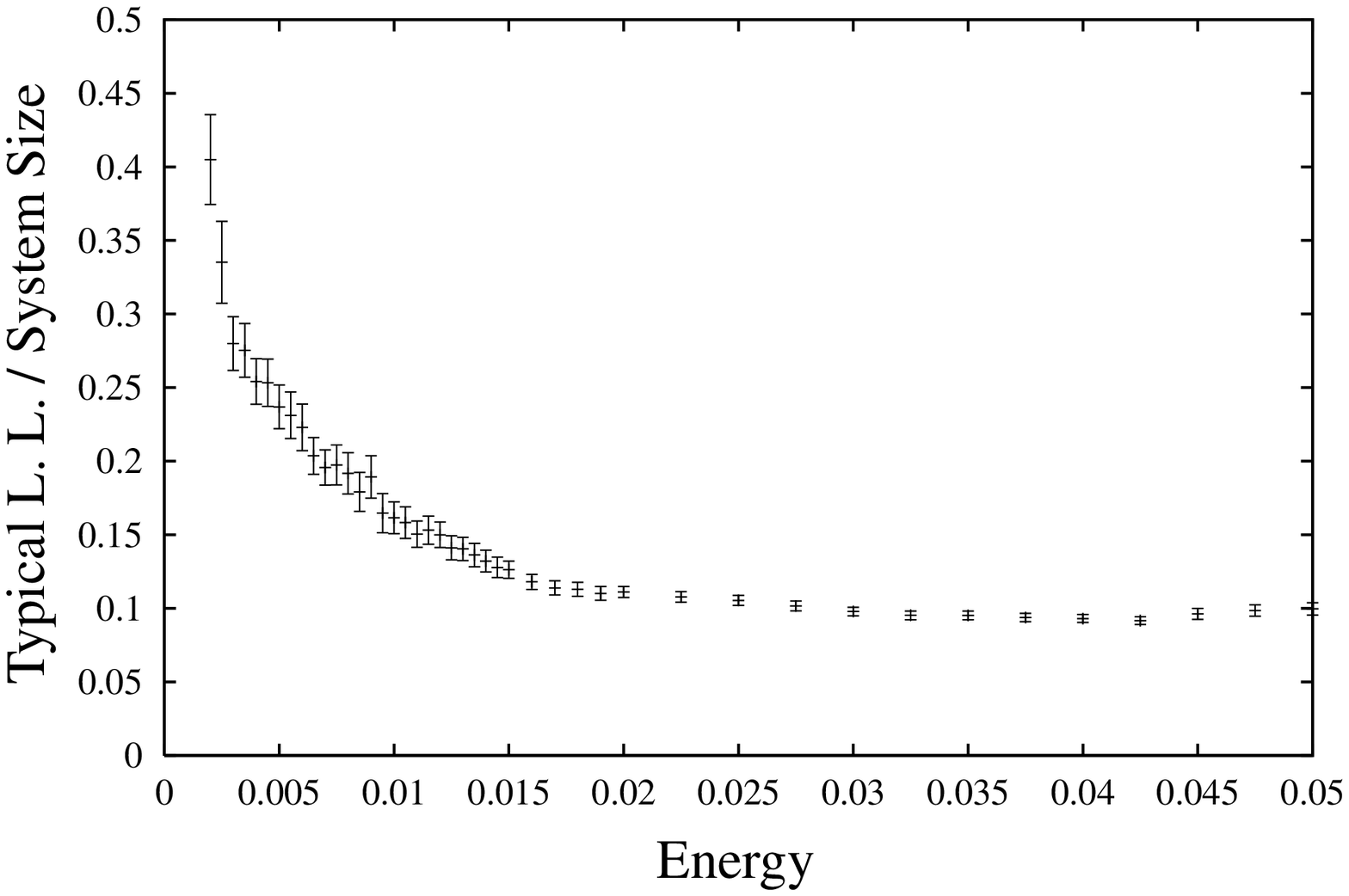}
}}
\put(-4,-3){
\centerline{
\epsfysize=3.8cm
\epsfbox{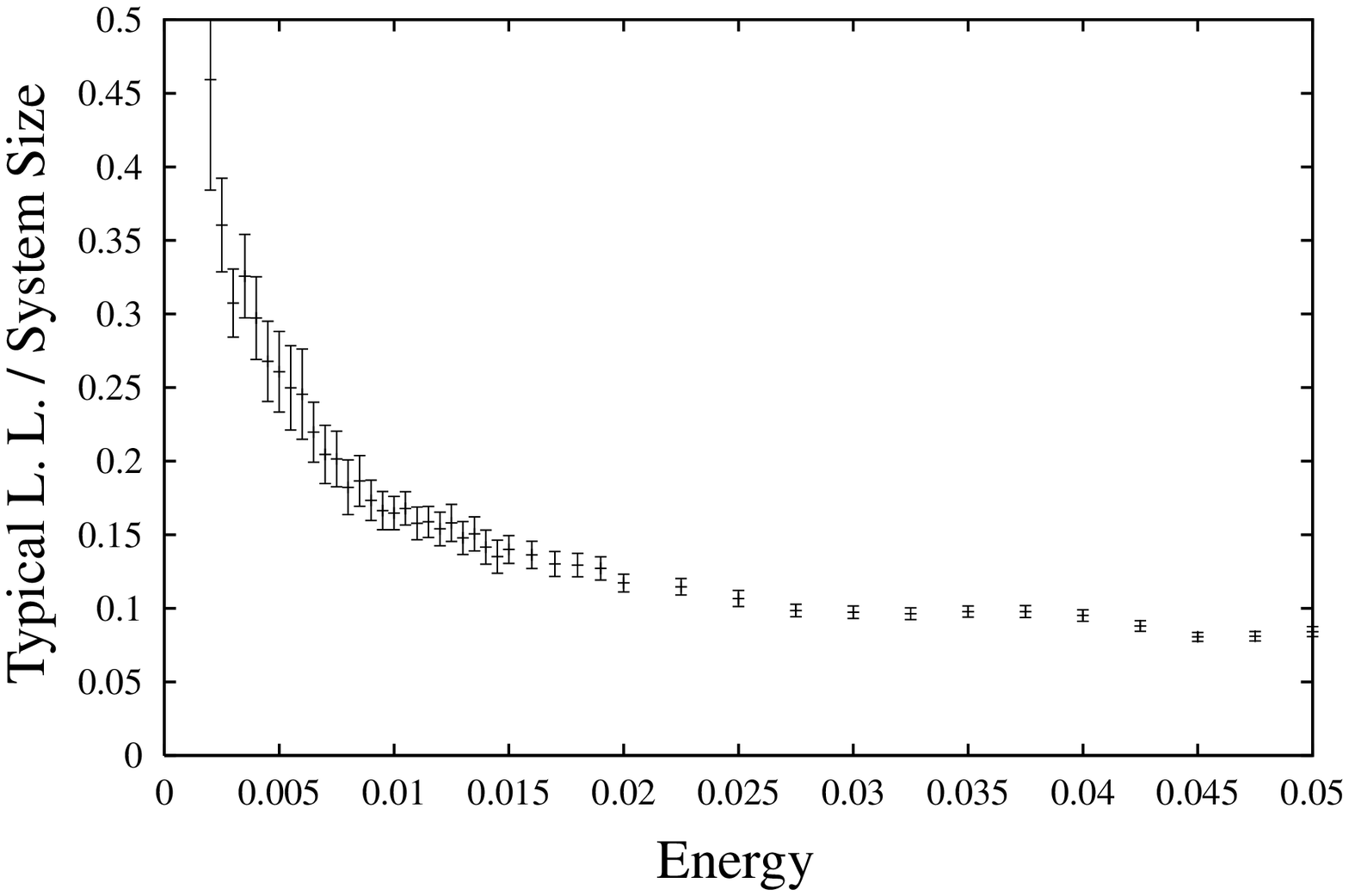}
}}
\end{picture}

\vspace{35mm}
\caption{ The energy dependence of localization length in the case of 
 power-law correlated random mass:  We set $L$(system size)$=50$ and
 $1000$ kinks in the system. $\alpha_{pw}$'s and $C$'s in
 Eq.(\ref{mmens}) are given in Table 2. Localization length
 is averaged within energy slice $\Delta E=0.003, 0.004, 0.006$ and $0.01$
 for $0<E<0.01, 0,01<E<0.015, 0.015<E<0.02$ and $0.02<E$ respectively. 
 The interval of data points are
 $5 \times 10^{-4}, 1 \times 10^{-3}$ and $2.5 \times 10^{-3}$ for
 $E<0.015, 0.015<E<0.02$ and $0.02<E$ respectively. 
 Divergences of localization
 length are observed only at $E=0$.}
\vspace{0cm}
\end{center}
\end{figure}

\begin{center}
\begin{tabular}{cccc}
 \cline{2-4} 
 {$$}&{$ $}&{$\alpha_{pw}$}&{$C$} \\ \cline{2-4}
 &\hspace{3mm} top left \hspace{3mm} & $0.68$\hspace{3mm} & $0.016$\hspace{3mm}\\  
 {Table 2. Parameters of power-law correlation.}\hspace{10mm}&\hspace{3mm}  top right\hspace{3mm} & $0.34$\hspace{3mm} & $0.056$\hspace{1mm}  \\  
 &\hspace{3mm} bottom\hspace{3mm} & $0.099$\hspace{3mm} & $3.1$\hspace{3mm} \\
 \cline{2-4}
 \end{tabular}
\end{center}

In Fig.11, we show the energy dependences of the localization length
in the system with power-law correlated random mass.
We choose value of the parameter $\alpha_{pw}$ as in Table 2.
These results show that localization length diverges only at
$E=0$, as in the case of white-noise and exponentially correlated disorder.
This behaviour does not depend on the value of $\alpha_{pw}$.

We also study the system size dependence of the localization length.
Results are shown in Fig.12.
We conclude that extended states exist only at the band center $E=0$.

\begin{figure}
\label{fig:powerlaw3}
\begin{center}
\unitlength=1cm

\begin{picture}(18,3.2)
\centerline{
\epsfysize=4cm
\epsfbox{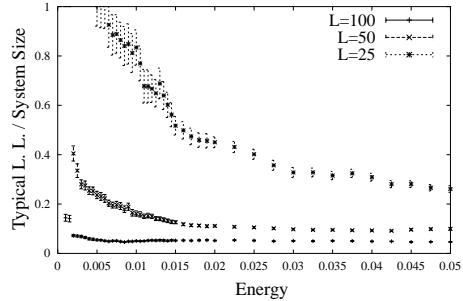}
}
\end{picture}
\vspace{-0.3cm}
\caption{ The system size dependence of typical localization length in 
the case of 
 power-law correlation: Parameters except for the system size and number
 of kinks are the
 same used in the top right figure
 in Fig.11. (Here we fix the number of kinks per system size.)
 Divergence of localization length is observed at $E=0$ in each
 case.}
\vspace{-0.5cm}
\end{center}
\end{figure}

\begin{figure}
\begin{center}
\unitlength=1cm
\begin{picture}(10,4.5)
\put(-8,1){
\centerline{
\epsfysize=3.5cm
\epsfbox{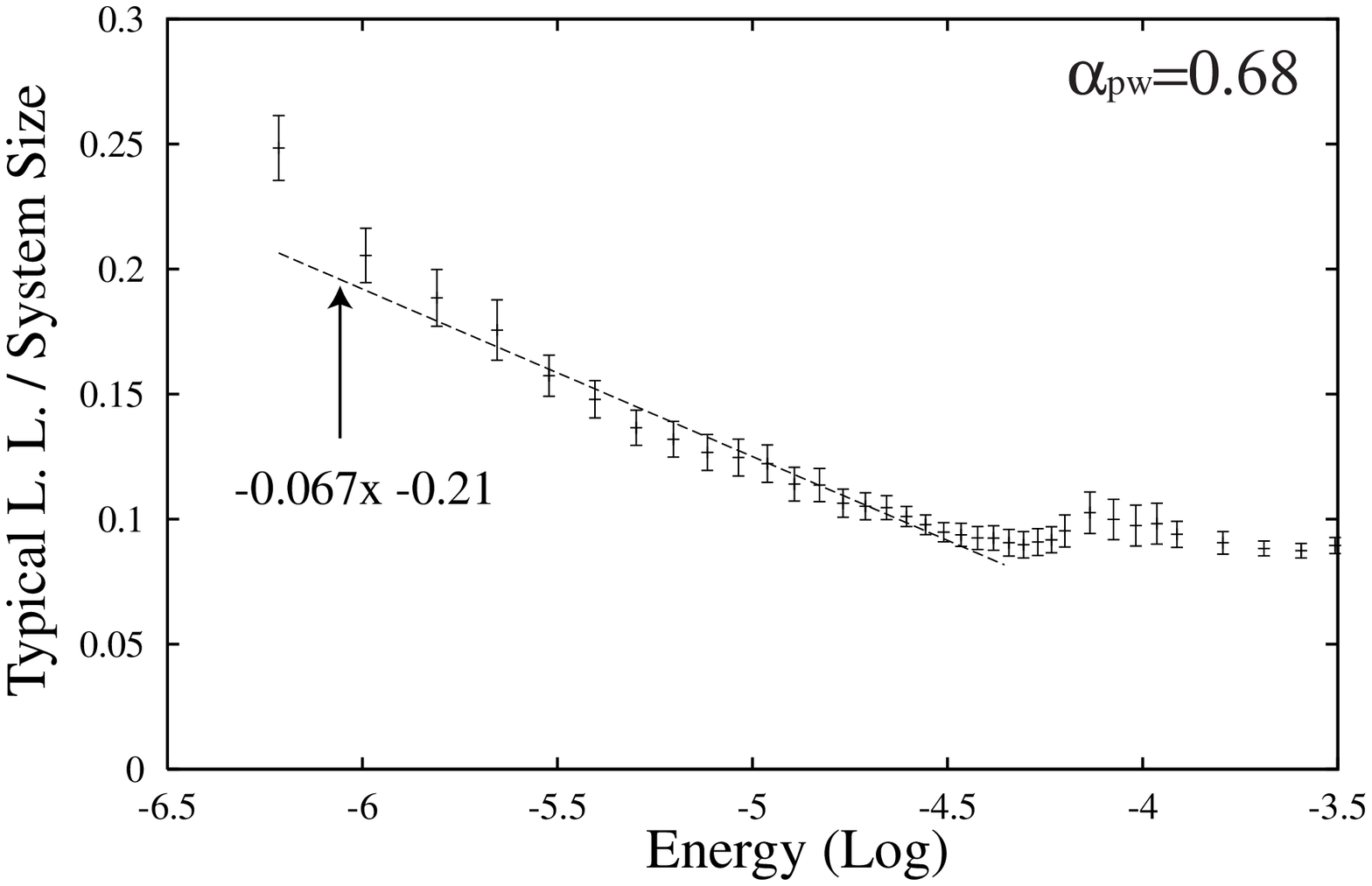}
}}
\put(-1,1){
\centerline{
\epsfysize=3.5cm
\epsfbox{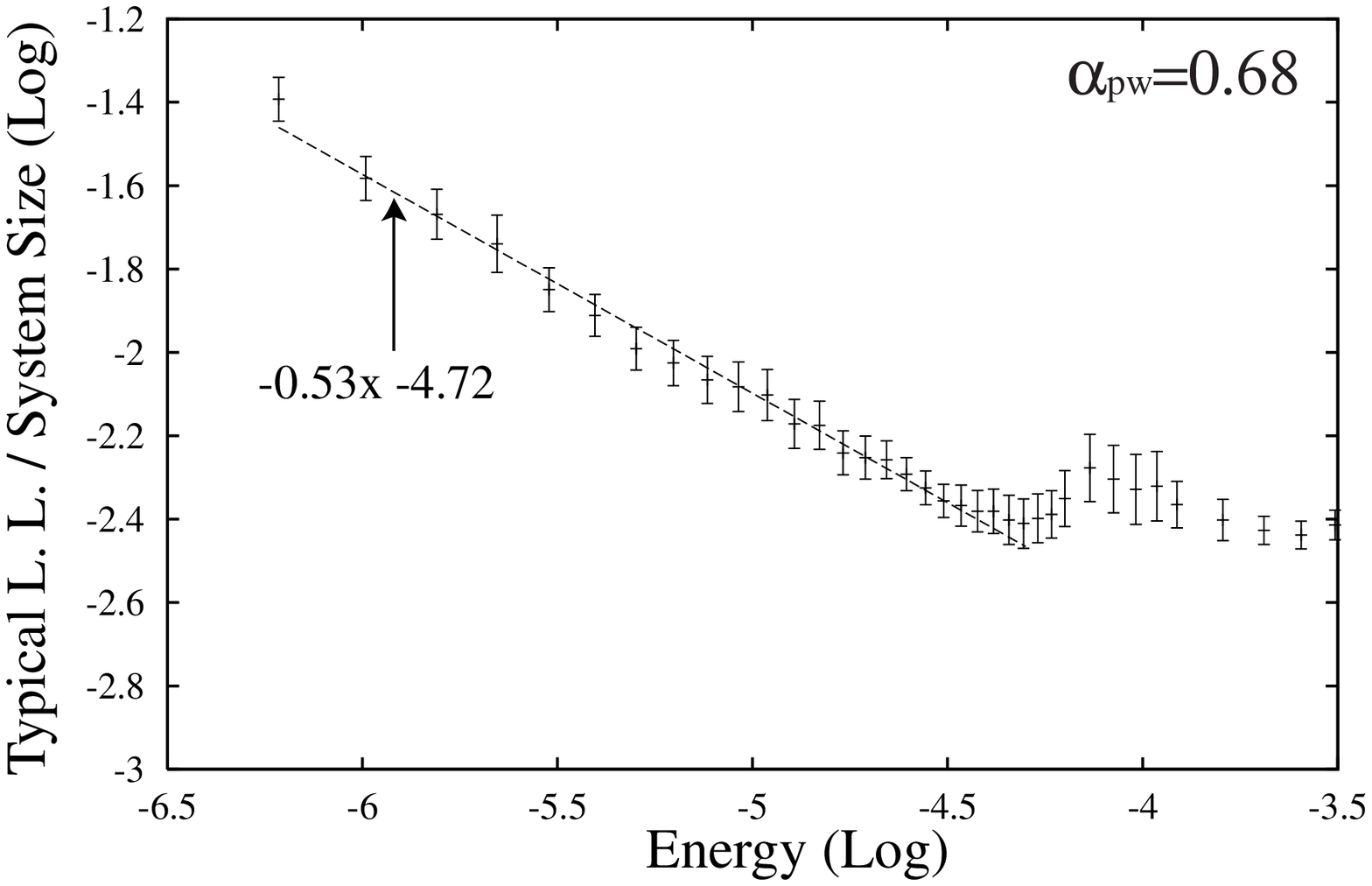}
}}
\put(-8,-2.5){
\centerline{
\epsfysize=3.5cm
\epsfbox{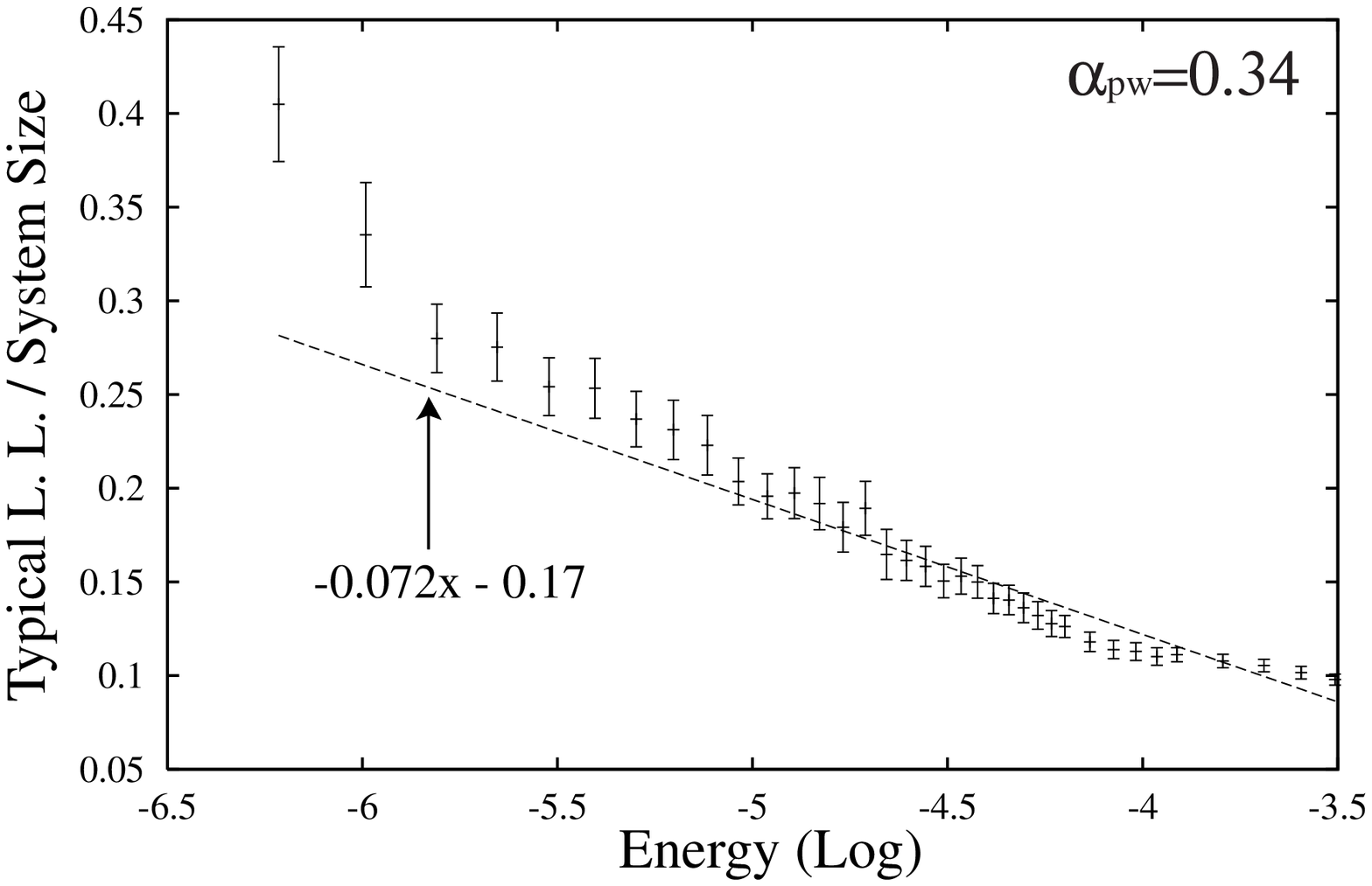}
}}
\put(-1,-2.5){
\centerline{
\epsfysize=3.5cm
\epsfbox{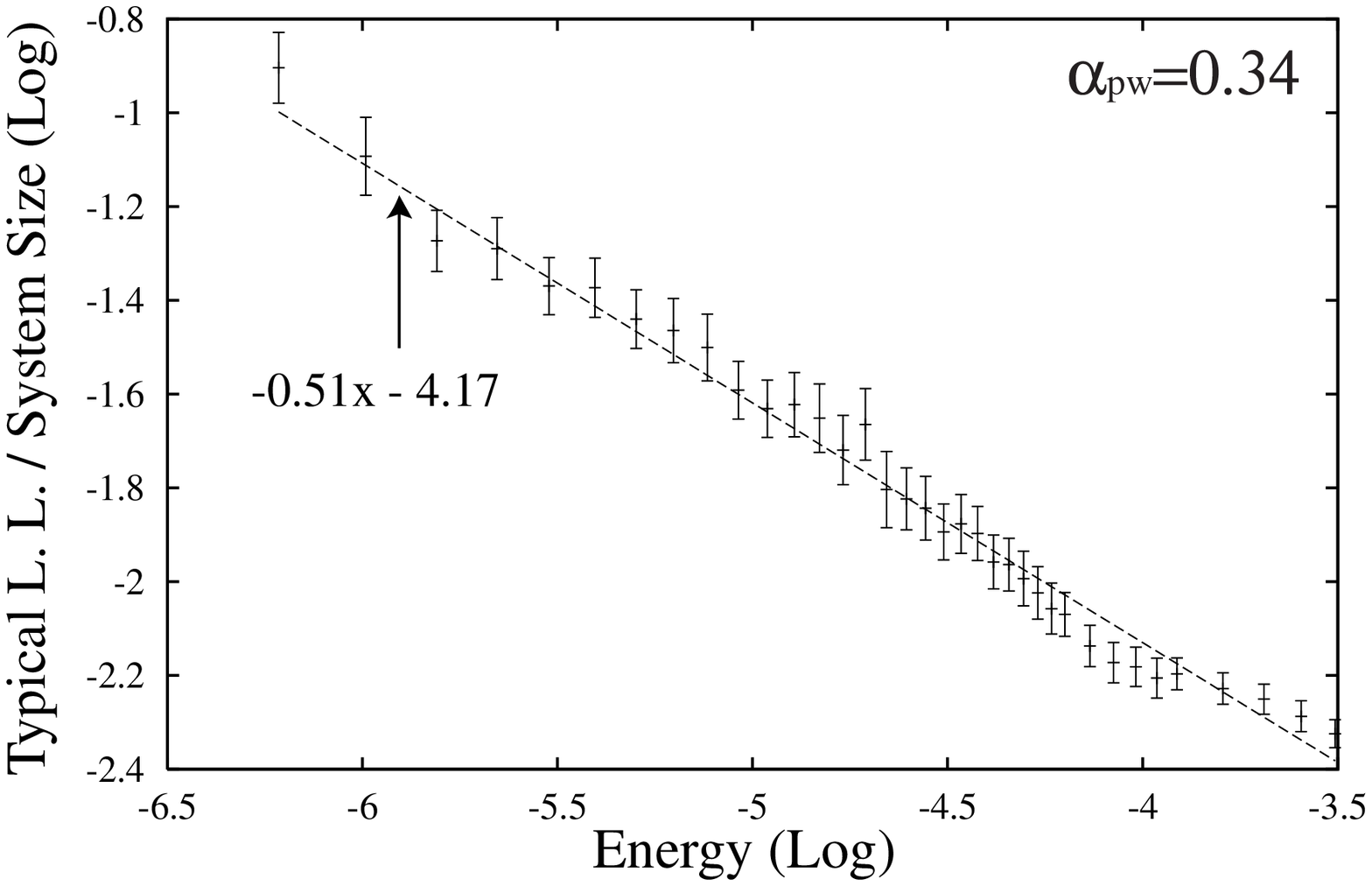}
}}
\put(-8,-6){
\centerline{
\epsfysize=3.5cm
\epsfbox{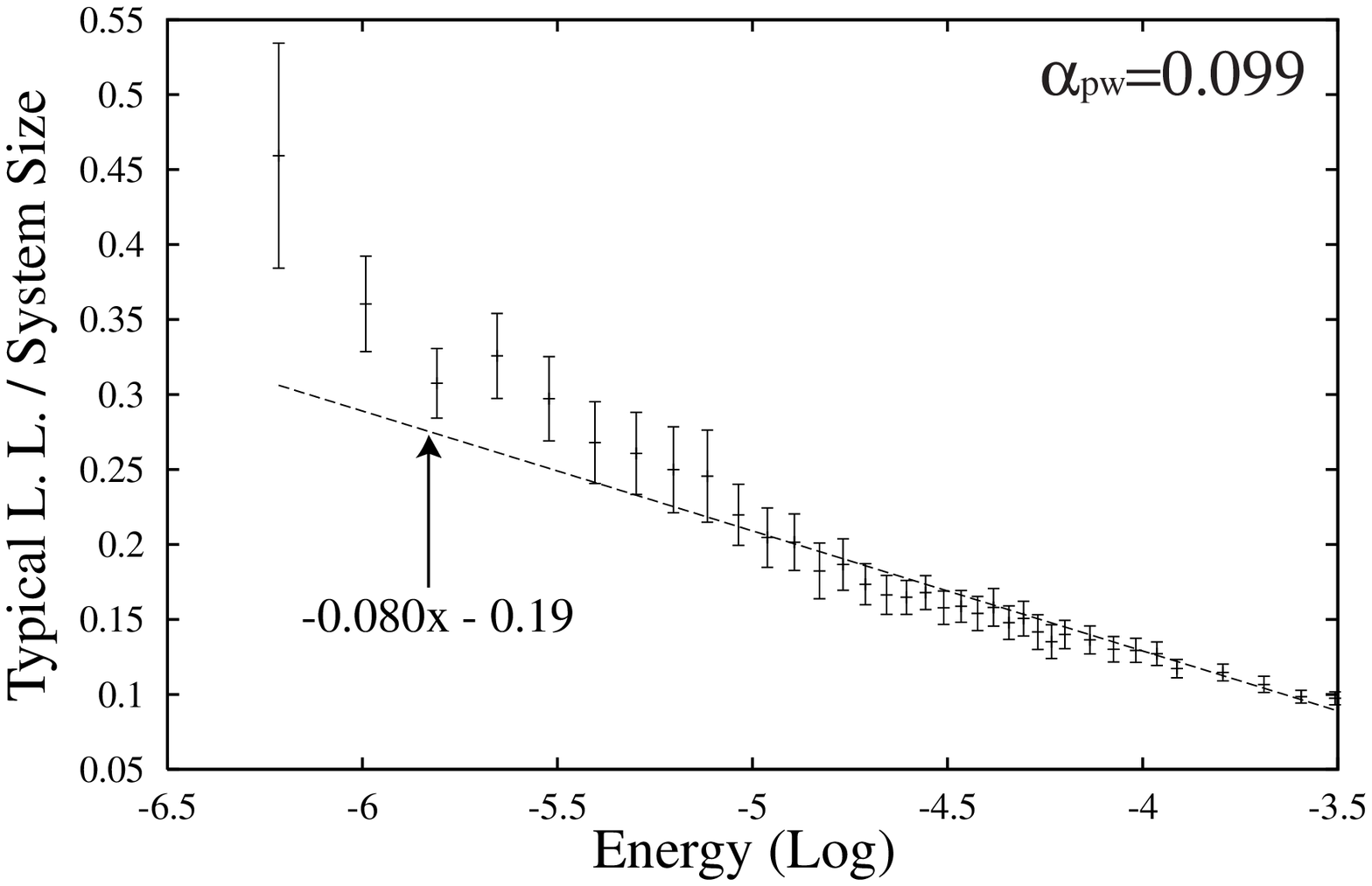}
}}
\put(-1,-6){
\centerline{
\epsfysize=3.5cm
\epsfbox{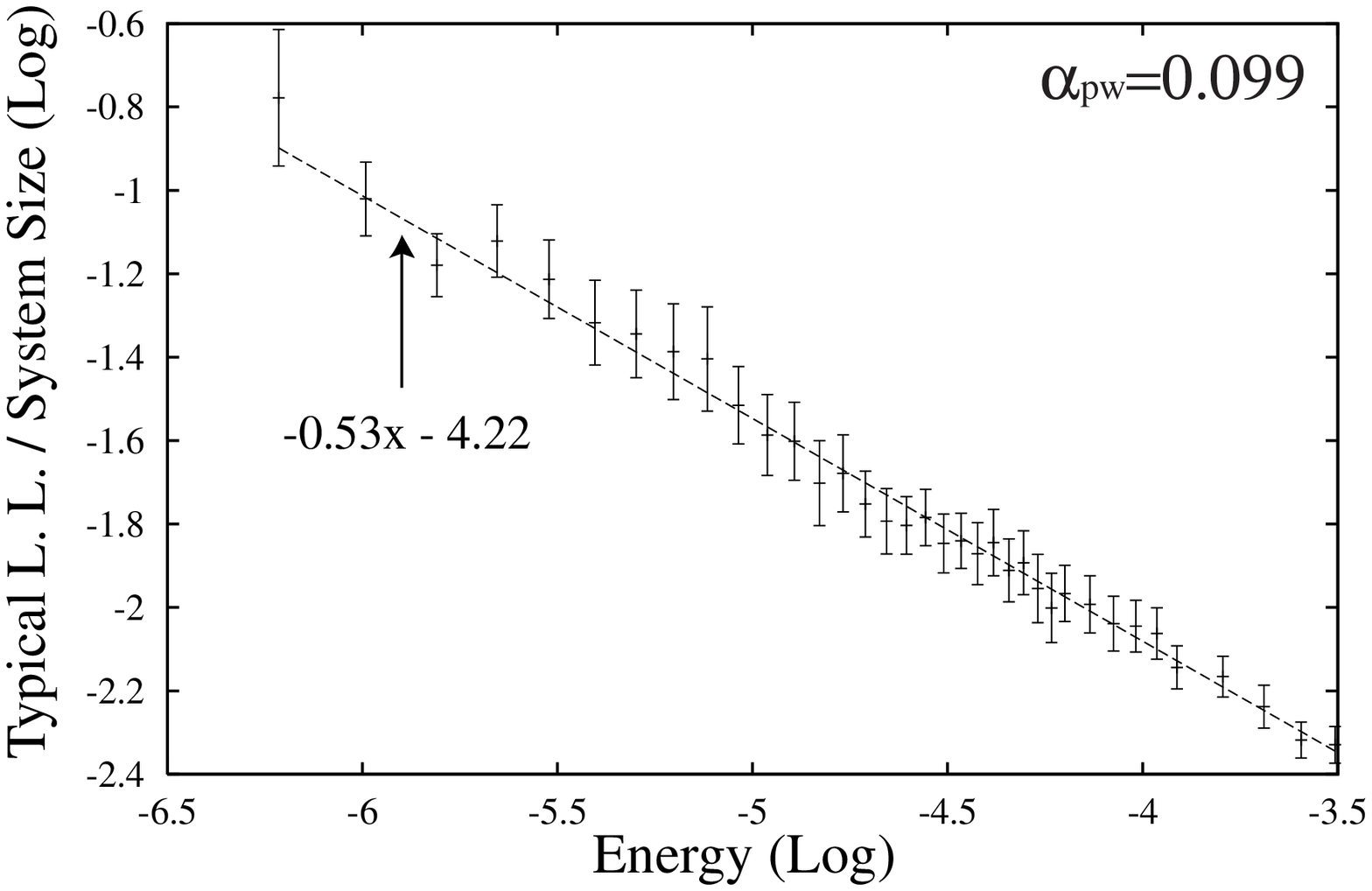}
}}
\end{picture}
\end{center}
\vspace{6cm}
\caption{Log-uni and log-log plot of the data shown in Fig.9: The 
 left column is log-uni and the right one is log-log plot.}
\label{fig:powerlaw2}
\vspace{0cm}
\end{figure}

\begin{center}
\begin{tabular}{cccc} \cline{1-4} 
 {$\alpha_{pw}$}&\hspace{3mm}{incline} \hspace{3mm}
 & \multicolumn{2}{c}{$\chi^{2}$/freedom} \\
 { }&{(log-log)}\hspace{3mm}&{(log-log)}&{(log-uni)} \\ \cline{1-4}
 \hspace{3mm} $0.68$\hspace{5mm} & $-0.53$\hspace{5mm}& $0.25$ & $1.08$ \\  
 \hspace{3mm} $0.34$\hspace{5mm} & $-0.51$\hspace{5mm}& $0.83$ & $3.05$ \\  
 \hspace{3mm} $0.099$\hspace{5mm} & $-0.53$\hspace{5mm}& $0.22$ & $1.10$ \\ \cline{1-4}
 \end{tabular}
 \end{center}
\vspace{0mm}
{\small Table 3. Results of linear fit to log-uni and log-log plot:
 We use the data in the range $-6<\log E<-3.5$ 
 ($-6<\log E<-4.4$ for $\alpha_{pw}=0.68$). The incline in the table
 corresponds to $-\beta$ in Eq.(\ref{Epower}). $\chi^{2}$ values for
 log-log plot are estimated smaller than log-uni plot.} 

In order to investigate the singular behaviours of the Dirac fermions with
the long-range correlated random mass,
we change the linear scale of energy and localization length in Fig.11
to log-uni and log-log scale plots in Fig.13.

If the data in the log-uni scale are on a straight line,
the energy dependence of localization length $\xi(E)$ is given by
\begin{equation}
\xi(E) \propto |\log E| + \mbox{const.} \hspace{5mm}(E \rightarrow 0),
\end{equation}
as in the white-noise case.
On the other side, if the data in the log-log scale are on a straight line,
the energy dependence of $\xi(E)$ is given as Eq.(\ref{Epower})
$\xi(E)\propto E^{-\beta}$ up to possible $|\log E|$ corrections.
From Fig.13 and Table 3, we can conclude that $E$-dependence of $\xi(E)$ is 
given by Eq.(\ref{Epower}). 
The exponent $\beta$ is estimated as in Table 3.

The above singular behaviour in the present system at $E=0$, i.e.,
Eq.(\ref{Epower})
is also confirmed by the calculation of the DOS near the band center $E=0$.
We show the energy dependence of the DOS in Fig.14. 

\begin{figure}
\label{fig:powerlaw4}
\begin{center}
\unitlength=1cm
\begin{picture}(17,4.5)
\put(-4,1){
\centerline{
\epsfysize=3.7cm
\epsfbox{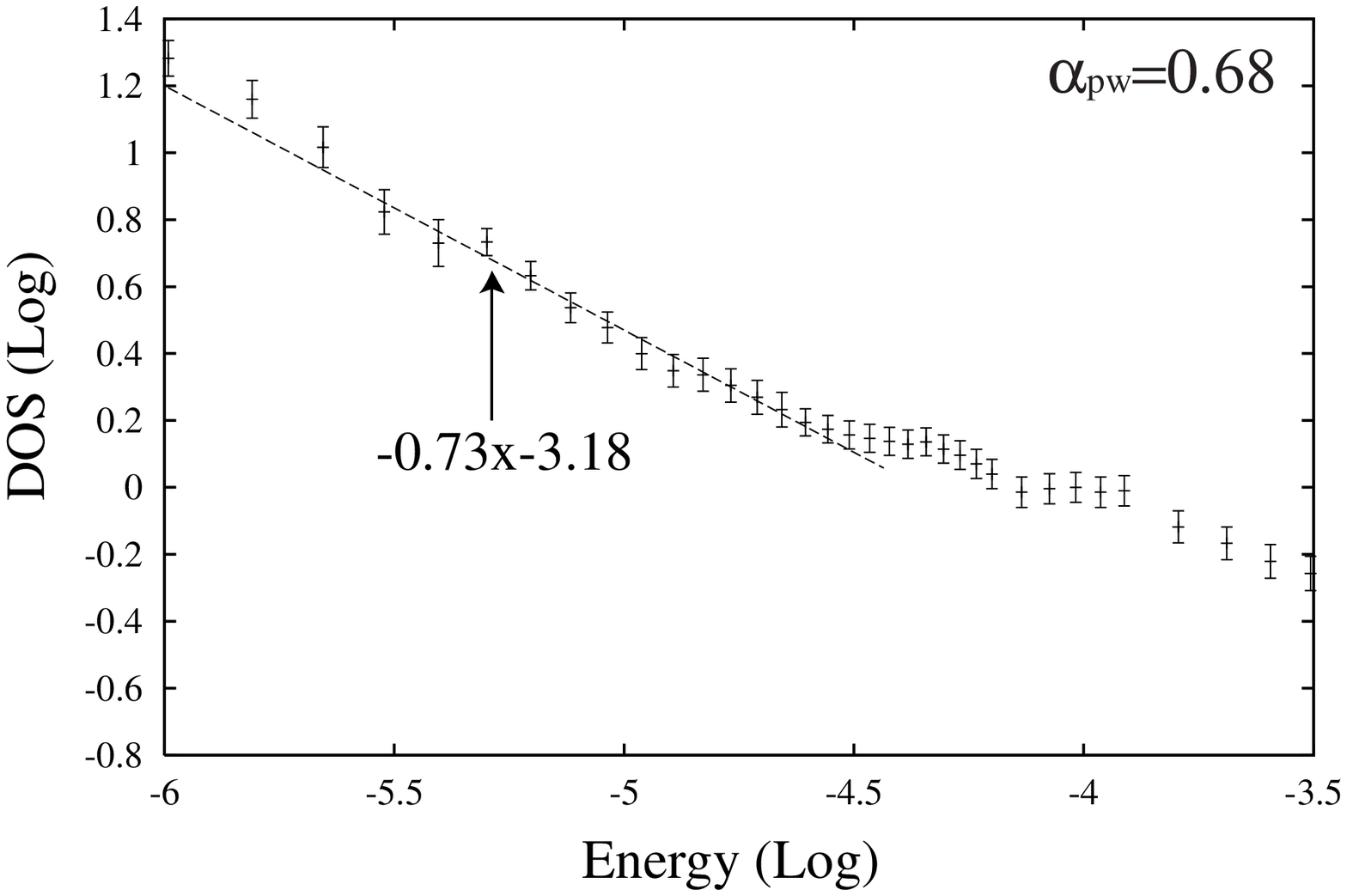}
}}
\put(3,1){
\centerline{
\epsfysize=3.7cm
\epsfbox{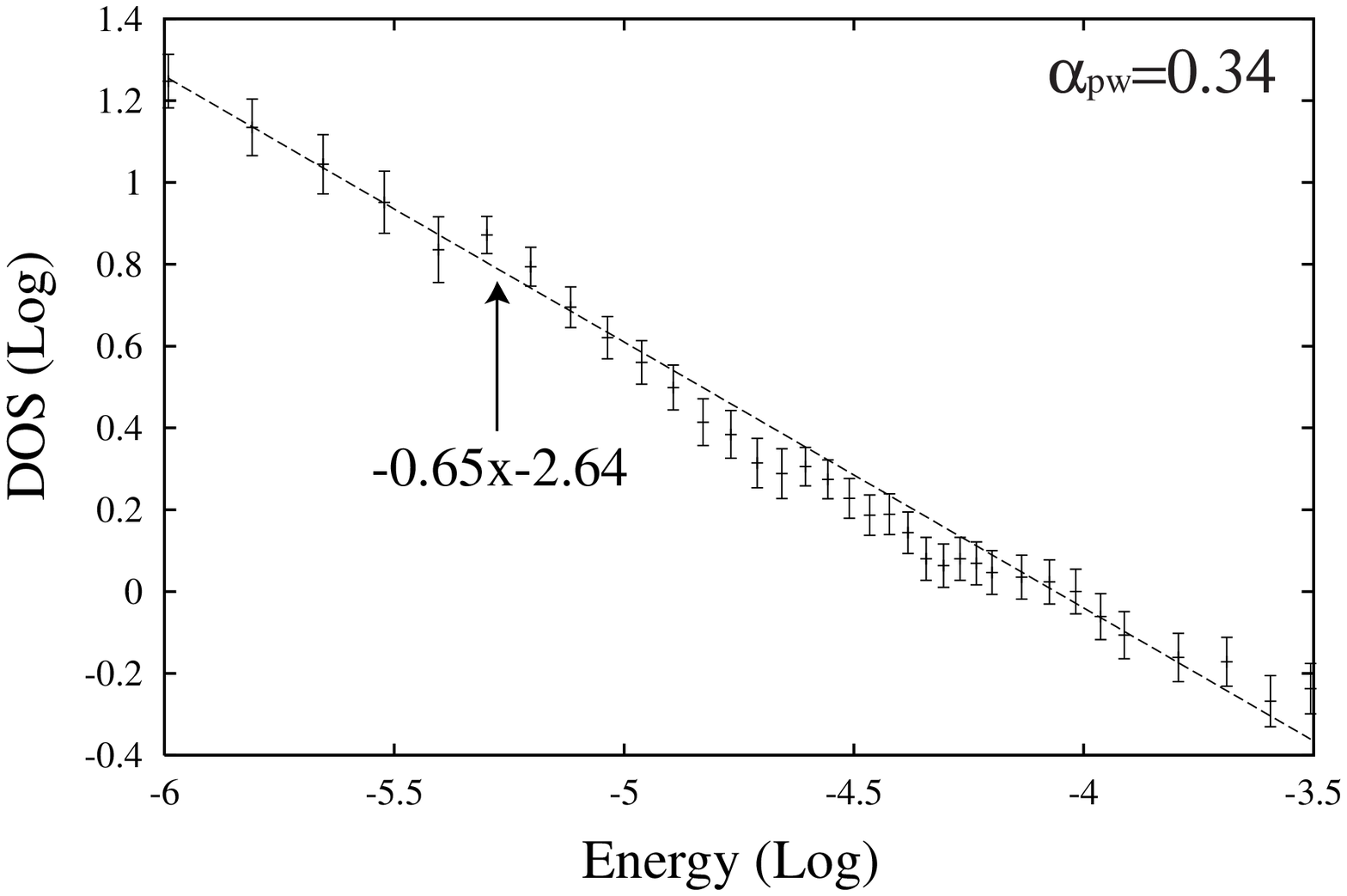}
}}
\put(-4,-3){
\centerline{
\epsfysize=3.7cm
\epsfbox{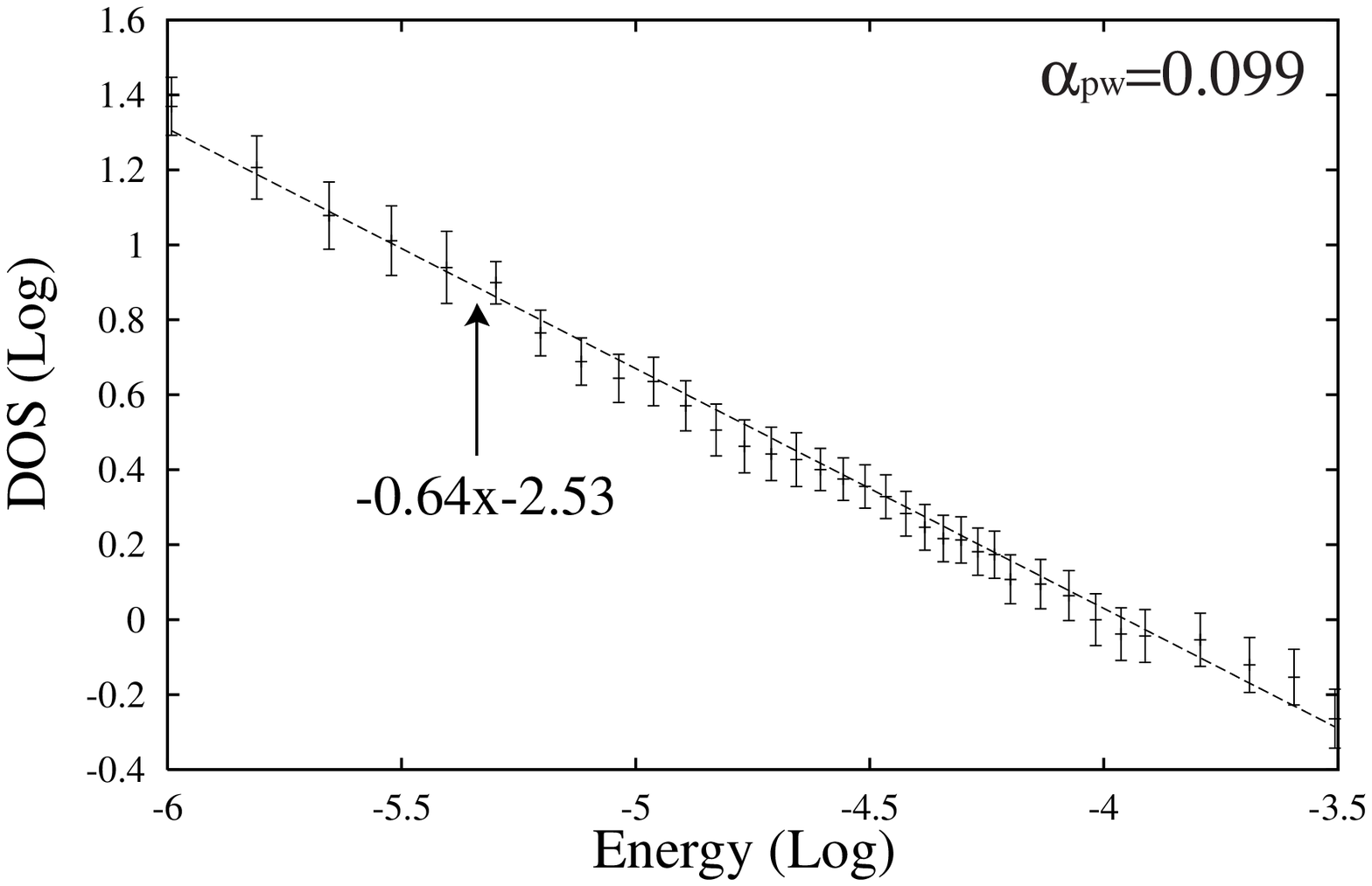}
}}
\end{picture}

\vspace{35mm}
\caption{ The energy dependence of DOS in the case of 
 power-law correlated random mass:
 DOS is calculated as the number of states within energy slice
 $\Delta E=0.002, 0.006$ and $0.01$ for $E<0.005, 0.005<E<0.01$
 and $0.01<E$ respectively.
 The interval of data points are
 $5 \times 10^{-4}, 1 \times 10^{-3}$ and $2.5 \times 10^{-3}$ for
 $E<0.015, 0.015<E<0.02$ and $0.02<E$ respectively. 
 DOS are normalized at $\log E=-4.0$.}
\vspace{-0.5cm}
\end{center}
\end{figure}

\begin{center}
\vspace{2mm}
\begin{tabular}{ccc} \cline{1-3} 
{$\alpha_{pw}$}&\hspace{3mm}{incline} \hspace{3mm}
 & {$\chi^{2}$/freedom} \\ \cline{1-3}
 \hspace{3mm} $0.68$\hspace{5mm} & $-0.73$\hspace{5mm}& $0.86$ \\  
 \hspace{3mm} $0.34$\hspace{5mm} & $-0.65$\hspace{5mm}& $0.97$ \\  
 \hspace{3mm} $0.099$\hspace{5mm} & $-0.64$\hspace{5mm}& $0.25$ \\ \cline{1-3}
 \end{tabular}
\end{center}

{\small Table 4. Results of linear fit to DOS log-log plot:
 We use the data in the range $-6<\log E<-3.5$ $(-6<\log E<-4.4$ for
 $\alpha_{pw}=0.68)$.
 The incline in the table corresponds to $-\beta'$
 in Eq.(\ref{Epower3})}.
\vspace{8mm}

The result indicates that the DOS $\rho(E)$ diverges as
 \begin{equation}
 \label{Epower3}
 \rho(E) \propto E^{-\beta'} \hspace{5mm}  (E \rightarrow 0),
 \end{equation}   
with the constant $\beta'$ given in Table 4. 
From the Thouless formula
 Eq.(\ref{eq:thouless}), $\beta$ in Eq.(\ref{Epower}) and
 $\beta'$ in Eq.(\ref{Epower3}) must satisfy the relation like
 \begin{equation}
 \label{relbeta}
 \beta+\beta'=1.
 \end{equation}
 From Tables 3 and 4, it is seen that $\beta + \beta'$ is almost unity
as it is expected.
In order to determine the exponent $\gamma$ of
the possible $|\log E|$ correction in Eqs.(\ref{rhoE}) and (\ref{xiE}),
calculation of the localization and the DOS for wider range of $E$ is required.

In the above we concluded that the power-law correlation of the random mass
in Eq.(\ref{mmens}) does not influence the phase structure of the system and
extended states appear only at $E=0$.
This result may seem to be in contradiction to the results obtained in
 the previous papers \cite{lyra,izrailev,IzKrUl}, in
which nonlocally-correlated Anderson, random hopping and 
random Kronig-Penney (random delta potential) models are studied and
in each model
the presence of a {\em nontrivial mobility edge} is observed at a finite
energy.
However, the correlations of randomness in their models are different 
from those of the present study.
Actually in one of these papers \cite{lyra},
 the following correlation was used for the
one-dimensional random hopping tight binding(RHTB) model,
\begin{equation}
\label{eq:ML}
[ (\epsilon_{i} - \epsilon_{j})^{2} ]_{ens} \propto |i-j|^{\omega -1},
\end{equation}
where $\omega$ is a non-negative constant and  
they concluded that a delocalized phase appears in a finite region
around $E=0$ in the case of $\omega=1.5$. 
As the random-mass Dirac fermion is a low-energy effective field theory of the 
RHTB model, the above result is relevant to the present study.
As in the present study we consider the case $\omega <1$, their results
are {\em consistent} with ours.

We use the random variables in Eq.(\ref{eq:ML}) for $m(x)$ in the
present field-theory model and calculate the localization length as before. 
The result of the numerical study is shown in  Fig.15. 
From these calculations, we cannot find any {\em finite region}
of extended states around 
$E=0$ even in the case $\omega=1.5$. 
However, the system size may not be large enough for this calculation.
 
\begin{figure}
\label{fig:ML}
\begin{center}
\unitlength=1cm

\begin{picture}(17,5)
\put(-4,1){
\centerline{
\epsfysize=4cm
\epsfbox{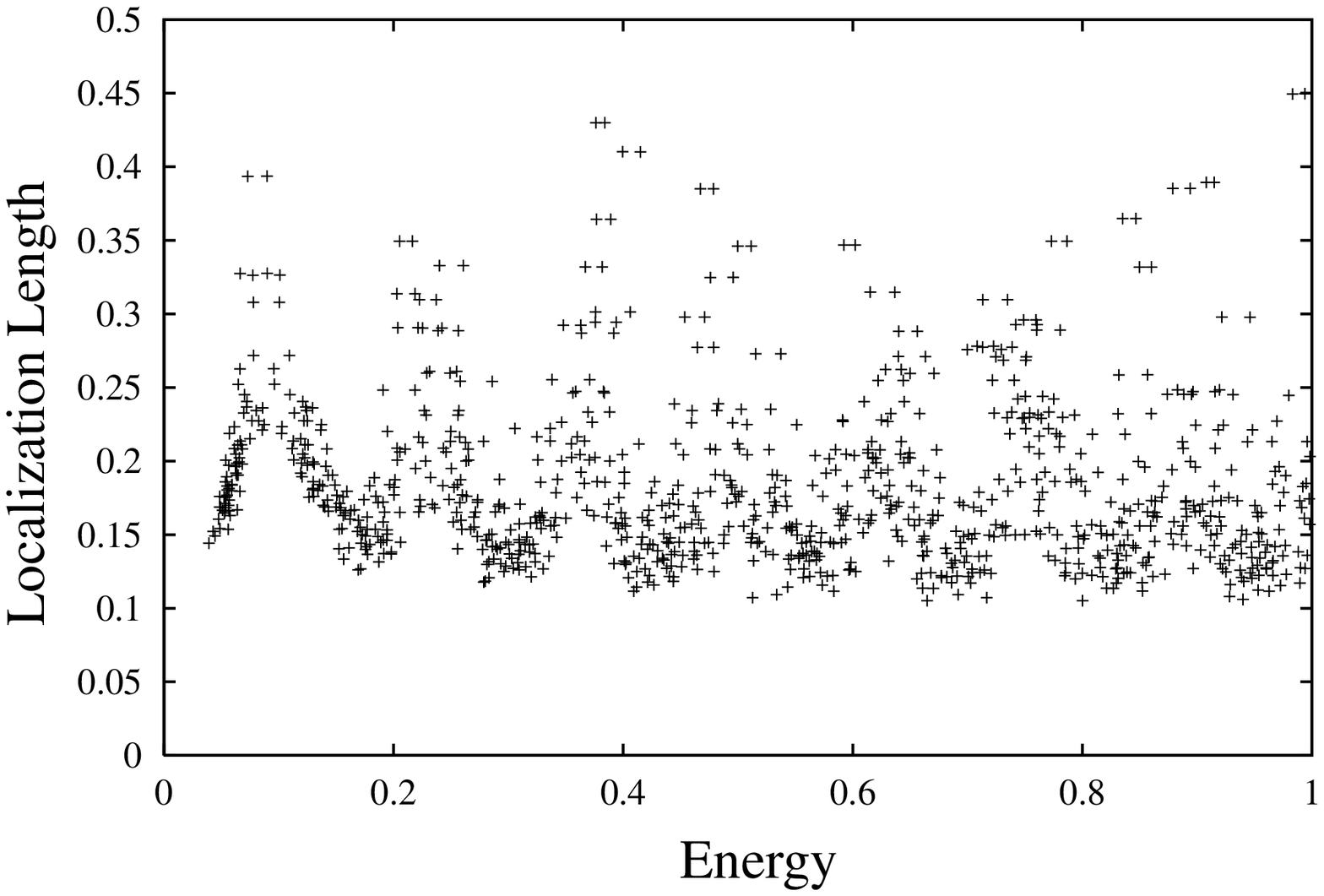}
}}
\put(3,1){
\centerline{
\epsfysize=4cm
\epsfbox{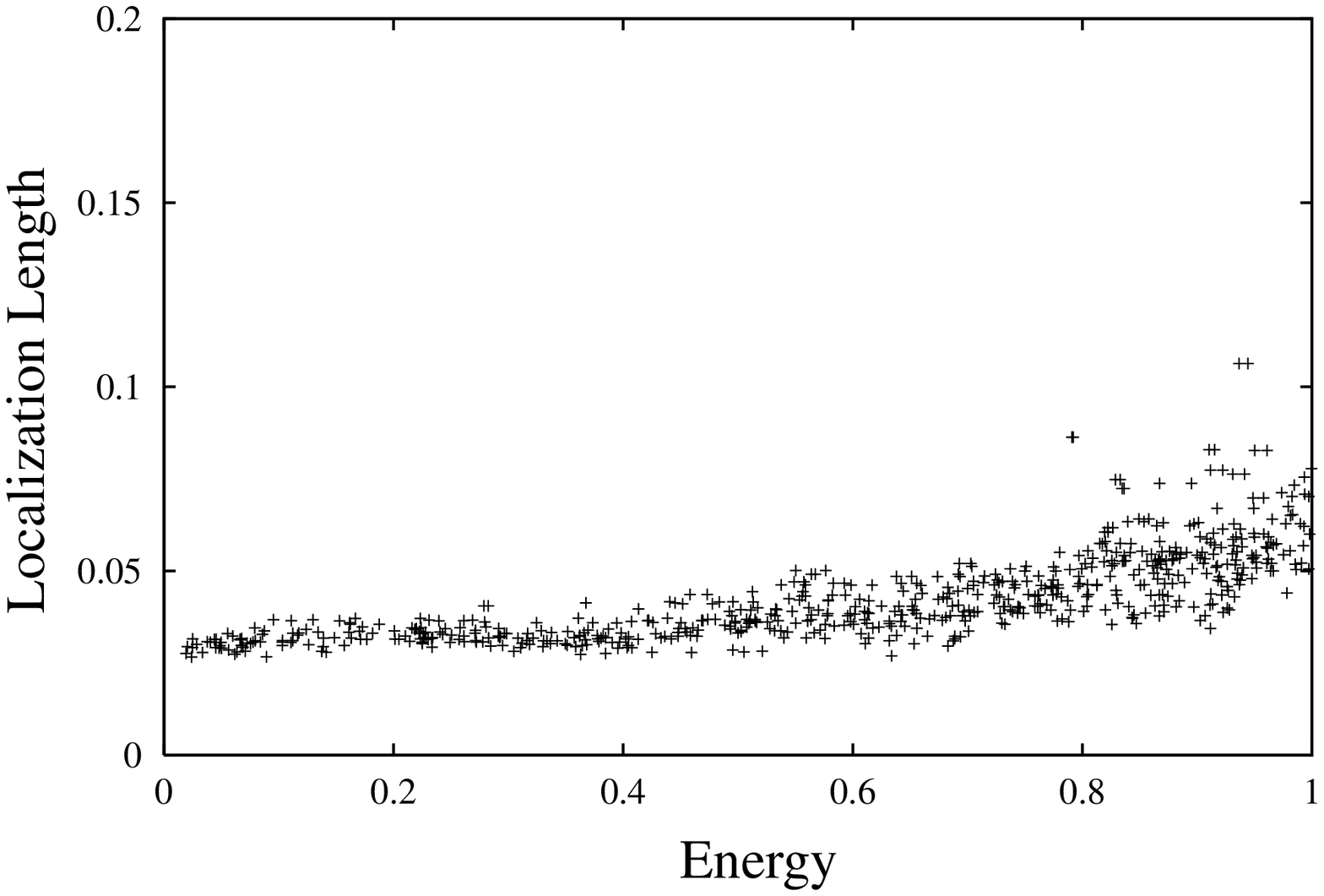}
}}
\end{picture}

\caption{ Localization length and energy of the eigenstates 
 in correlated disorder system: Correlation of $m(x)$ is given in
 Eq.(\ref{eq:ML}). 
 We set L(system size)=50 and 200 kinks in the system.
 We set $\omega=0$ at the left and $\omega=1.5$ at the right.}
\end{center}
\end{figure}

\section{Conclusion}

In this paper, we studied the Dirac fermion with the long-range
correlated random mass by using the TMM and IVPM.
We are especially interested in the delocalization transition or
the existence of a finite mobility edge.
Recent studies on the Anderson model, random hopping model
 and random Kronig-Penney model
indicate the existence of finite mobility edges.
However we found that extended states exist only at the band
center $E=0$ as in the white-noise random variable case.

We also studied the localization length as a function of $E$
rather in detail.
We found that in the case of short-range correlation 
$\xi(E) \propto |\log E|$ for small $E$ whereas
$\xi(E) \propto E^{-\beta}$ $(\beta \sim 0.5)$ for the
long-range correlation of the random mass.
The above conclusion is supported by the calculation of the DOS
through the Thouless formula.
We hope that the above behaviour of the localization length is
observed by experiment of the random spin model because the
localization length is directly related with the spin-spin
correlation length there. 

We can also calculate the multi-fractal scaling indices
directly from wave functions of the random-mass Dirac fermions.
Values of the indices may depend on the decay power of the correlation
of the random mass.
Results will be reported in a future publication \cite{KI2}.

\clearpage


\begin{thebibliography}{99}
\bibitem[*]{Takeda} Electronic address: takeda@icrr.u-tokyo.ac.jp
\bibitem[**]{Ichinose} Electronic address: ikuo@ks.kyy.nitech.ac.jp

\bibitem{PPs}K. Takeda, T. Tsurumaru, I. Ichinose, and M. Kimura,
Nucl. Phys. B {\bf 556} (1999) 545.

\bibitem{KI1}K. Takeda and I. Ichinose, J. Phys. Soc. Japan {\bf 70}
 (2001) 3623.

\bibitem{SUSY}L. Balents and M. P. A. Fisher, Phys. Rev. B {\bf 56}
 (1997) 12970. 

\bibitem{IK}I. Ichinose and M. Kimura, Nucl. Phys. B {\bf 554} (1999) 607; 
ibid, B {\bf 554} (1999) 627.

\bibitem{lyra}F. A. B. F. de Moura and M. L. Lyra, Phys. Rev. Lett. {\bf 81}
(1998) 3735; Physica A {\bf 266} (1999) 465.

\bibitem{izrailev}F. M. Izrailev and A. A. Krokhin, Phys. Rev. Lett. {\bf 82}
(1999) 4062.

\bibitem{IzKrUl} F. M. Izrailev, A. A. Krokhin and S. E. Ulloa, 
 Phys. Rev. B {\bf 63} (2001) 041102.

\bibitem{HN}N. Hatano and D. R. Nelson,
Phys. Rev. Lett. {\bf 77} (1996) 570; Phys. Rev. B {\bf 56} (1997) 8651.

\bibitem{herbut}I. F. Herbut, cond-mat/0007266.

\bibitem{comment}J. W. Kantelhardt et. al.
Phys. Rev. Lett. {\bf 84} (2000) 198; F. A. B. F. de Moura and M. L. Lyra,
Phys. Rev. Lett. {\bf 84} (2000) 199.

\bibitem{thouless}D. J. Thouless, J. Phys. C {\bf 5} (1972) 77;
 G. Theodorou and M. H. Cohen, Phys. Rev. B {\bf 13} (1976) 4597.

\bibitem{KI2}K. Takeda and I. Ichinose, work in progress.

\end{thebibliography}
\end{document}